\documentclass{emulateapj}

\def\simg{\mathrel{\rlap{\raise 0.511ex \hbox{$>$}}{\lower 0.511ex \hbox{$\sim$}}}}
\def\siml{\mathrel{\rlap{\raise 0.511ex \hbox{$<$}}{\lower 0.511ex \hbox{$\sim$}}}}

\def\eps{\varepsilon}
\def\Ep{E_b} \def\Ec{E_c} \def\Ei{E_i} 
\def\Ip{I_b} \def\Fp{F_b}
\def\F0310k{{\cal F}(0.3-10\,{\rm k})}
\def\bL{\beta_L} \def\bH{\beta_H} 
\def\bx{\beta_{\rm x}} \def\bo{\beta_o}
\def\ag{\alpha_\gamma} \def\at{\alpha_t}
\def\ap{\alpha_p} \def\app{\alpha_{pp}}
\def\Da{\Delta \alpha}
\def\Dagrb{\Delta \alpha_{grb}}
\def\Daag{\Delta \alpha_{aglow}}
 
\def\tej{t_{ej}} \def\to{t_o} \def\tg{t_\gamma}
\def\tp{t_p} \def\tpp{t_{pp}}
\def\calD{{\cal D}}
\def\Gc{\Gamma_c}
\def\thc{\theta_c}
\def\thobs{\theta_{obs}}
\def\Gctc{\Gamma_c \theta_c}
\def\Gt{\Gamma\theta}

\def\psicr{\psi'_{cr}}
\def\chinu{\chi^2_\nu}
\def\deg{{\rm o}}

\def\lra{\longrightarrow}

\def\ti{t_i}             
\def\tang{t_{ang}}       
\def\td{t_{dyn}}         
\def\tce{t_c}            

\def\h{\hspace*{-1mm}}
\def\hh{\hspace*{-2mm}}
\def\hhh{\hspace*{-3mm}}

\def\Meszaros{M\'esz\'aros~}

\begin{document}

\parskip 5pt
\topmargin 1cm

\title{Determination of the Angular Distributions of Dynamical and Emission Parameters of GRB Relativistic Outflows}

\author{A. Panaitescu}

\affil{Intelligence and Space Research, MS D440, Los Alamos National Laboratory, Los Alamos, NM 87545, USA}

\begin{abstract}

 A previously-developed analytical treatment of the {\sl Larger-Angle Emission} model for GRBs and afterglows is further developed 
to include the effect of a fixed magnetic field orientation. This model 
has the uncanny ability to unify the prompt and delayed emissions and to account easily for the entire diversity of Swift X-ray light-curves.

 The $LAE$ model is applied to seven GRBs (060526, 060614, 060714, 060729, 061007, 061121, 100902, 130427A) with well-sampled 0.3 keV/10 keV 
Swift/XRT and optical light-curves, displaying well-monitored four phases (GRB, Tail, Plateau, post-Plateau), to illustrate its power and limitations 
and to determine its five fundamental parameters. Tight constraints are not always obtained from fitting even the best light-curves available.

 The one-and-only Core parameter $\Gctc$ is determined by the GRB and afterglow temporal milestones (i.e. by the dynamical range of the Tail 
and Plateau), and may have an universal value around 3.0, with narrower Cores (smaller $\thc$) being more relativistic (higher $\Gc$).
 Flat Plateaus require an Envelope angular structure close to $\Gamma(\theta) \sim \theta^{-2}$.

 A magnetic field parallel to the direction of the relativistic flow yields more curved prompt/GRB and delayed/afterglow light-curves 
than a perpendicular or isotropic field, with curvature $\Da$ defined as the difference between the flux asymptotic power-law decay indices
at the beginning and end of each phase. 
 The curvature $\Dagrb$ of the GRB light-curves arising from the {\sl uniform Core} is set only by the $B$-field orientation, 
and a perpendicular or isotropic $B$-field is equally often found, but a parallel field is never identified for the Core.
 The curvature $\Daag$ of the afterglow light-curves from the {\sl power-law Envelope} depends on $B$ and three other parameters. 
The Envelope $B$-field orientation can be identified for only three afterglows, where a parallel $B$ is found. 

{\sl Unified Astronomy Thesaurus Concepts}: High-Energy Astrophysics, Non-thermal Radiation Sources, Gamma-Ray Bursts, Relativistic Jets, 
  Theoretical Models

\end{abstract}


\vspace*{3mm}
\section{\bf Introduction}

 In the last several years, the Larger-Angle Emission ({\bf LAE}) ("large" would be a misnomer for angles of just a few degrees) 
released during the GRB phase from the fluid moving at angles larger than $\Gamma^{-1}$ relative to the observer's central line-of-sight 
has emerged as an intriguing alternative (G. Oganesyan et al 2020) to the {\sl relativistic blast-wave} model (P. \Meszaros \& M. Rees 1997), 
the leading model for the afterglow emission over the last three decades (but that model still "rules", and with good reason).

 The dissipation mechanism and radiation process producing the $LAE$ are those for the GRB emission and no specific identification 
is made for the former, thus the fundamental properties of the $LAE$ (peak-flux, peak-energy, spectral energy distribution -- SED)
are not calculated explicitly. The only requirement is that the SED is a broken power-law, which points to a non-thermal radiation process
(synchrotron) from a plasma which is optically-thin at the photon energies of interest (optical and X-ray).

 Lacking a specific dissipation mechanism implies that all the physics of the $LAE$ model lies in its treatment of the differential radiation
relativistic beaming across the radiating surface. In other other words, the $LAE$ model is just a prescription for including the relativistic 
enhancement of the radiation from a surface endowed with an angular structure of both its local Lorentz factor and comoving-frame emission 
characteristics.
 However, the non-uniformity of the Lorentz angular distribution is all that is needed for $LAE$ to unify the GRB/prompt and the afterglow/delayed 
emissions, and to account for the temporal diversity of all Swift/XRT GRB+afterglow data, but that success stops at the diversity of 
differential light-curve behaviors at various photon energies (optical and X-ray).

 The $LAE$ model is a "single-source" model, thus it can account only for light-curves (at various photon-energies) that are well-coupled,
i.e. with similar flux-decay rates and synchronous flux-decay breaks.
 A decoupling of light-curves at two photon energies can be acquired if one or two spectral breaks separate those photon energies, 
which are predominantly optical and X-ray. 

 The passage of a spectral break through the observing band leads to a chromatic light-curve break, confined only to that photon-energy, 
but none of the possible breaks (the "injection break" where the freshly-injected/uncooled electrons radiate, and the "cooling break" of 
where radiate electrons that cool on a dynamical timescale) can account for the timing and magnitude of the X-ray light-curve breaks at 
the end of the X-ray light-curve Plateau. 

 Therefore, {\sl chromatic X-ray light-curve Plateaus} (A. Panaitescu et al 2006a) cannot be accounted by the $LAE$ model. 
If X-ray afterglows with chromatic light-curve breaks arise from the $LAE$, then their optical emission must originate somewhere else: 
the relativistic blast-wave. A hybrid origin of such afterglows could naturally occur because the $LAE$, which peaks in the hard X-rays 
during the prompt phase, could be brighter than the external shock in the X-rays at all times, but always dimmer in the optical.

 In constrast, the {\sl relativistic blast-wave model}, being a two-source model (a reverse-shock into the incoming GRB ejecta and 
a forward-shock in the circumburst medium), can explain alone the entire optical and X-ray emissions of afterglows with both coupled 
(same flux-decay rates, achromatic/synchronous light-curve breaks) and decoupled light-curves (different flux-decay rates, chromatic breaks), 
as shown by F. Genet, F. Daigne \& R. Mochkovitch (2007) and by Z. Uhm \& A. Beloborodov (2007).

\vspace*{1mm}
\subsection{\bf Unification of prompt and delayed emissions}

 The $LAE$ model unifies the GRB and afterglow emissions in a simple way, relying only on the differential relativistic beaming on the emitting 
surface, provided that that surface has a Lorentz factor that decreases with angle from the outflow's symmetry axis (G. Oganesyan et al 2020), 
as one might expect for a jet interacting with an in-falling stellar envelope. 

 The blast-wave model unifies the GRB and afterglow X-ray emissions by attributing them to the reverse-shock (Z. Uhm \& A. Beloborodov 2007).
That choice is motivated by the reverse-shock emission being more "flexible" than the forward-shock's, which is the result of the reverse-shock 
dynamics and number of ejecta uncooled electrons being more flexible than for the forward-shock. 

 However, it is possible to assign
the entire GRB+afterglow emissions to the forward-shock if the relativistic outflow is a "naked" uniform Core (without an Envelope) 
that produces hard X-ray emission during the GRB phase and soft X-ray emission (and below) during the afterglow phase owing to a re-energized 
forward-shock, in order to account for the X-ray afterglow light-curve Plateau. That re-energization could be acquired 
1. by means of delayed ejecta produced by a long-lived accretion process (P. Kumar, R. Narayan \& J. Johnson 2008; J. Cannizzo, E. Troja \& N. Gehrels 2011),
2. by instantaneously-ejected ejecta having a range of initial Lorentz factors (M. Rees \& P. \Meszaros 1998), or 
3. by the wind driven by a magnetar (B. Zhang \& P. \Meszaros 2001).

\vspace*{1mm}
\subsection{\bf Afterglow diversity} 

 The most complex X-ray afterglows display {\bf four phases}: GRB prompt, GRB Tail, afterglow Plateau, afterglow post-Plateau: \\
 1. During the {\bf GRB prompt} phase (partially shown in some Figures), the hard X-ray flux may have large fluctuations which are not captured 
by the $LAE$ model because its assumption of an instantaneous emission makes it relevant only for the end of the GRB prompt phase, while its
assumption of a uniform Core leads to a steady light-curve. \\
 2. The {\bf GRB Tail} occurs over a dynamical time-range of about 1 dex, but that is dependent on the epoch of instantaneous emission 
(see Figures). The hard X-ray flux decay ranges between $t^{-2}$ to $t^{-4}$, which is also dependent on the emission epoch. \\
 3. The {\bf afterglow Plateau} starts at 100-1000s after trigger, lasts for 1-2 dex in time and is characterized by a $t^{-0.3 \pm 0.3}$ flux decay. \\
 4. Finally, the {\bf afterglow post-Plateau} starts at 5-50 ks after trigger and displays a "normal" $t^{-1.0 \pm 0.4}$ flux decay. \\
 Some fraction of Swift afterglows display three phases (GRB prompt without a Tail, followed by an afterglow without a Plateau), or only 
two phases (GRB prompt followed by a single power-law decaying afterglow). 
 
 The $LAE$ model can easily account for the diversity of GRBs and their afterglows (with two, three, or four light-curve phases) by simply 
adjusting the one-and-only Core $\Gctc$ parameter (Lorentz factor times Core opening) and the exponent of the power-law structure of the 
Envelope Lorentz-factor angular distribution (A. Panaitescu 2020), with the off-axis observer location being another important source of 
light-curve diversity (S. Ascenzi et al 2020).

 The reverse-shock model can also achieve that goal easily by adjusting the Lorentz factor and density of the incoming ejecta, due to that
the X-ray emission is produced by the high-energy electrons freshly accelerated at the reverse-shock. That correlates the X-ray flux 
to those two ejecta properties. 

 The forward-shock model could explain the observed afterglow diversity if the Core could be as narrow as the inverse of its Lorentz factor 
(which leads to a GRB without a Tail) and if energy-injection could be sometimes weak or negligible (which leads to an afterglow without a Plateau, 
displaying a fast flux-decay).

\vspace*{1mm}
\subsection{\bf GRB/afterglow modeling and testing}

 Numerical {\sl fits} (with very little physics) using the $LAE$ have been employed for the prompt and delayed, optical and X-ray observations 
in only a few cases (G. Oganesyan et al 2020, A. Panaitescu 2025). 
 During the last 2.5 decades, numerical {\sl modelings} (with more physics) using {\sl analytical} treatments the blast-wave 
dynamics were done for a few dozen radio, optical, and X-ray afterglow light-curves, but using only the forward-shock emission. 
 The latter is also true for fits employing 1D hydrodynamical {\sl simulations} that track the propagation of strong, relativistic 
shocks over distances much larger than the thickness of the shocked fluid (A. Panaitescu \& P. \Meszaros 1998) or for fits with relativistic 
jets whose emission is calculated using analytical prescriptions for the dynamics of 2D jets (G. Ryan et al 2020), based on interpolations 
of the results of 2D hydrodynamical simulations (H. van Eerten, A. van der Horst \& A. MacFadyen 2012), with the extra restriction of a homogeneous 
circumburst medium (CBM).
(wind-like CBMs, as expected for the massive stellar progenitors of long GRBs, are included in analytical models for the jet dynamics)

 To a large extent, both the $LAE$ and the blast-wave models are untestable, as they both have enough freedom to produce almost any 
afterglow light-curve. 
 In its simplest form, the $LAE$ model has one Envelope dynamical parameter (for the outflow Lorentz factor) and two "emission characteristics" 
parameters (for the break-energy and luminosity at that energy), whose angular distributions (uniform or not) can yield various light-curve decays 
and features. 
 The blast-wave model has more free parameters: for the dynamics of reverse-shock (set by the Lorentz factor and mass radial distributions 
in the incoming ejecta), for the dynamics of the forward-shock (set by the blast-wave energy and ambient medium structure/density), 
plus two micro-physical shock parameters, all of which can also lead to very diverse light-curves. 

 Thus, testing these models is more a question of assessing if they reach their goals in a simple way, if their assumptions are natural, 
and if the best-fit parameters required to accommodate observations are reasonable, with a lot of uncertainity on what "natural" and 
"reasonable" mean. 
These evaluations can be qualitative, done by comparing light-curve features with analytical expectations, or quantitative, done through 
a comparison between model emission and the data, with the caveat that such evaluations test only the adopted model variant, with all 
its assumptions and short-cuts/approximations, which are often forgotten in the end.
 Thus, model testing beyond proof-of-concept may be a too lofty and too frustrating exercise.

\vspace*{1mm}
\subsection{\bf What is new here ?}

 In this work, we continue to develop the analytical formalism for $LAE$ light-curves (A. Panaitescu 2020, 2025) by including the effect 
of a unidirectional magnetic field, with emphasis on a parallel $B$ (oriented along the direction of flow) and a perpendicular $B$. 
Previous calculations of the $LAE$ were done ignoring the flux-correction factor associated with a specific $B$ orientation. 
Given that an isotropic $B$ also leads (naturally) to a time-independent flux-correction factor, it can be said that previous calculations 
are valid for an isotropic $B$.

 The analytical model is then used to fit the prompt/burst and delayed/afterglow optical and Swift/XRT X-ray light-curves of seven GRBs 
with well-coupled and well-monitored light-curves that display all four phases.
 The requirement of well-coupled light-curves is obvious: there is no point in applying a model where it is known to be insufficient.
 The requirement of well-monitored light-curves showing all four phases is imposed so that the number of observational constraints is maximal, 
with the hope of determining (tightly constrain) all fundamental model parameters, including the magnetic field orientation. The detection
of the GRB Tail and afterglow Plateau has the side benefit of ensuring an observer located close to the symmetry axis, as was assumed for 
the analytical $LAE$ model, because a significant off-axis observer locations would mix these two phases (S. Ascenzi et al 2020).

 To retain all the information in the XRT data, Swift 0.3-10 keV fluxes and effective spectral slopes are converted into specific fluxes 
at 0.3 keV and 10 keV, as described in A. Panaitescu (2020).
 The purpose of fitting GRB/afterglow light-curves at three photon-energies (optical, soft X-ray, mid-Xray) is to determine allowed ranges 
for the model parameters yielding best-fits.

\vspace*{2mm}
\section{\bf Analytical Formalism of the $LAE$ Model}

\vspace*{1mm}
\subsection{\bf Basic assumptions}

 The simple premise of the $LAE$ model is that the relativistic surface releases emission on a timescale shorter than
the spread in the observer-frame photon-arrival time $\tang = (r/c) (1 - \cos \delta \theta) = r/(2c\Gamma^2)$ 
(with $r$ the source radius, $\Gamma$ the outflow Lorentz factor), over the $\delta \theta = \Gamma^{-1}$ angular opening 
of the visible area (which is the region of maximal Doppler boost $D \simeq \Gamma$).

 This assumption is valid if all other timescales that determine the duration of emission at the photon energy $\eps$
of interest are smaller than $\tang$.

 One such timescale is the $\ti$ for injecting relativistic electrons in the region where they radiate, and the requirement
that $\ti < \tang$ means that electrons are injected over a duration shorter than the source age $r/c$ (or its dynamical
time $\td$) because, in the observer-frame, timescales are defined by the arrival-time of photons from a source moving
relativistically at $\Gamma$ toward the observer, thus $\td = (r/c)/(2\Gamma^2) = \tang$.
 This requirement rules out a long-lived shock origin for the burst, such as the forward-shock sweeping the ambient medium.

 For the remaining radiative cooling time $\tce$ of the electrons radiating synchrotron emission at energy $\eps$,
the condition $\tce < \tang$ means that electrons cool on less than the epoch of observation (measured from when the ouflow 
was ejected). For typically expected "conditions" in the GRB ejecta, the cooling timescale of electrons radiating at
$\eps = 100$ keV (during the burst) is of few tens of milliseconds. Then $\tce \sim \eps^{-1/2}$ for synchrotron emission
implies that electrons radiating at $\eps = 0.3-10$ keV (soft X-ray) cool on a subsecond timescale, while those radiating at
$\eps = 1-5$ eV (optical) cool over several seconds (during the afterglow). 
 Thus, the second requirement is always statisfied.
(The above requirements, $\ti, \tce < \td$, do not specify if electrons are cooled ($\tce < \ti$) or uncooled ($\ti < \tce$)).

 The great pay-off that follows from the above {\bf assumption $(i)$} {\sl of an instantaneous emission release} and from 
the assumptions of an on-axis observer and of an axially symmetric outflow (which imply that the outflow Lorentz factor
$\Gamma$ is a function of only $\theta$) is the one-to-one correspondence between the angle $\theta$ from where emission 
is released and the photon arrival-time $t$
\begin{equation}
  \left[\frac{\Gc}{\Gamma(\theta)}\right]^2 + (\Gc \theta)^2 = \frac{t}{\to} \;,\;
  \to = \frac{r}{2c\,\Gamma^2_c}
\label{pke}
\end{equation}
where {\bf $\Gc$ is the Core Lorentz-factor} and $\to$ is the observer-frame epoch when the instantaneuos emission is released 
(at radius $r$), {\sl measured from the epoch when the outflow was launched} (which can be after the GRB trigger time).

 For a specified $\Gamma(\theta)$ distribution, Equation (\ref{pke}) can be solved for $\theta(t)$, which enables an analytical formalism 
to be delevoped but, for that, further assumptions are needed: 
{\bf assumption $(ii)$} {\sl the outflow emission has axial symmetry} and 
{\bf assumption $(iii)$} {\sl the observer is located on that symmetry axis}. 

 Relaxing the former assumption is an interesting excercise but not immediately necessary/useful: existing observations should not require 
more model freedom because they do not set more stringent constraints than the number of model parameters (see below). Explaining X-ray 
flux fluctuations (flares) could benefit from a 2D outflow structure, but there is no obvious reason to suspect that a reverse-engineered 
azymuthal outflow structure would have any difficulties in accounting for flares. There may be some reward from the resulting 
best-fit parameters for the hot-spot responsible for the flare but here with are still working on the basic results for a 1D outflow.

 The latter assumption comes together with one related to the angular distribution of all relevant ejecta parameters: 
{\sl the outflow has a quasi-uniform Core} that is brighter than the Envelope, so that the outflow yields the usual morphology of 
a GRB light-curve, with a roughly constant (although variable) flux, followed by a steeply decay Tail. 
To account for the slow-decay/Plateau followed by a steeperly decaying post-Plateau (A. Panaitescu et al 2006b, B. Zhang et al 2006),
that Core should be surrounded by an Envelope where the Lorentz factor decreases outward sufficiently fast (as shown by G. Oganesyan 
et al 2020). For uniformity, we might call "assumptions" these last two requirements. 

 Then, the detection of a GRB favors observers located within the opening of the brighter Core because observers located outside it 
(in the dimmer Envelope) will not detect the dimmer burst (and its "orphan" afterglow). The assumption of Core uniformity is motivated 
by that the GRB flux, aside from fluctuations, is roughly flat, without an increasing or decreasing trend, as would be caused by 
a Core with a significant angular structure. Then, for a uniform Core, the assumption of an on-axis observer should have no important 
effect. Moreover, as discussed below, our use of GRBs with all four phases well-developed acts a selection effect for on-axis observer
locations. Conversely, if one aims to fit GRBs lacking Tail and Plateau, then off-axis, but in-Core, observer locations must be considered, 
and that has the extra benefit of constraining a dummy (but important) model parameter: the observer offset relative to the outflow symmetry axis.

 None of the above four assumptions are needed for an occasional numerical calculation of GRB/afterglow multi-wavelength light-curves,
but axial symmetry and on-axis observer location could be essential for sufficiently fast iterative searches of serious best-fits 
to real ligh-curves and reliable parameter determinations. Instead, those assumptions are essential for analytical results against 
which ones should always tests numerical results. 

 In summary, the major assumptions/approximations of this analytical treatment of the $LAE$ model are: \\
(1) an instantaneous emission release, \\
(2) an axially-symmetric outflow endowed with \\ 
(3) a Lorentz factor and emission characteristics that are uniform (angle-independent) in the outflow Core 
   and outward-decreasing with angle in the outflow Envelope (power-law is the best choice - \S\ref{temporal}, and \\
(4) an observer located on the jet symmetry-axis (which is not such a bad assumption as it seems at first - \S\ref{offaxis}).

\vspace*{1mm}
\subsection{\bf Light-Curve Temporal Milestones}
\label{temporal}

 The first term of Equation (\ref{pke}) is the contribution to the arrival-time $t$ from the outflow angular structure.
That term is dominant when $\Gt \ll 1$, thus the observer is within the cone of opening $\Gamma^{-1}$ of maximal relativistic 
enhancement (i.e. emission is beamed toward the observer). Although not evident at this time, the condition $\Gt < 1$ defines 
{\bf the GRB prompt and the afterglow post-Plateau}. In the latter case, Equation (\ref{pke}) leads to
\begin{equation}
 ({\rm post-Plateau}) \quad \Gamma(t) = \Gc \left( \frac{t}{\to} \right)^{-1/2} 
\label{Gpp}
\end{equation}
for {\sl any angular structure} $\Gamma(\theta)$.

 The second term of Equation (\ref{pke}) is the contribution to the arrival-time $t$ from the spherical curvature of the outflow.
That term is dominant when $\Gt \gg 1$, meaning that the observer is outside the cone of maximal relativistic boost
(i.e. emission is beamed away from the observer). Condition $\Gt > 1$ defines {\bf the GRB Tail and the afterglow Plateau}. 
In those cases, Equation (\ref{pke}) leads to 
\begin{equation}
\hh ({\rm Tail/Plateau}) \;\; \theta(t) = \Gc^{-1} \left( \frac{t}{\to} \right)^{\h 1/2} \hh = 
                                          \thc \left( \frac{t}{\tp} \right)^{\h-1/2} 
\label{ttp}
\end{equation}
for {\sl any angular structure} $\Gamma(\theta)$. 

 However, general results stop here and, to derive flux light-curves, $\theta(t)$ and $\Gamma(t)$ are required, thus a specific 
angular structure $\Gamma(\theta)$ must be considered. Numerical simulations of relativistic jets have yielded various results,
depending on the assumed initial jet structure, subsequent injection of energy, and interaction with the nearby environment:
from {\sl top-hat} (fig 21, 22) and {\sl Gaussian} (fig 27) $\Gamma(\theta)$ jet structures in M. Aloy, H. Janka \& E. Muller (2005), 
to {\sl single power-law} profiles in A. Mizuta \& M. Alloy (2009) (their fig 5), P. Duffell \& A. MacFadyen (2013) (fig 3), 
and D. Lazzati et al (2017) (fig 3), and to {\sl multiple power-law structures} (double or triple) in A. Tchekhovskoy, J. McKinney,
\& R. Narayan (2008) (fig 11).

 In the following derivations, we will assume that the relativistic outflow is endowed with a {\bf uniform Core of angular size $\thc$} 
and a {\bf power-law Envelope}
\begin{equation}
 \Gamma(\theta < \thc) = \Gc \;, \quad 
 \Gamma(\thc < \theta) = \Gc \left( \frac{\theta}{\thc} \right)^{-g}.
\label{Gplaw}
\end{equation}
 This {\bf assumption} $(o)$ of a {\sl power-law angular structure} is the most important/restrictive assumption and is justified 
by the need of a simple function with a minimal number of free parameters that could be determined by modeling real light-curves.
That condition is also satisfied by a generalized exponential $\Gamma = \Gc \exp\{-(\theta/\thc)^n\}$, which includes 
exponentials ($n=1$) and Gaussians ($n=2$), but those angular structures cannot account for afterglow Plateaus longer than 1.5 dex 
(dynamical range), as shown by G. Oganesyan et al (2020) for a Gaussian and by A. Panaitescu (2025) for any $n > 1$.

 Solving Equation (\ref{pke}) for the power-law structure of Equation (\ref{Gplaw}), leads to the following:
\begin{equation}
 ({\rm GRB/Tail}) \quad \theta(t) = \Gc^{-1} \left( \frac{t}{\to} - 1 \right)^{1/2} \quad \left[\Gamma(t) = \Gc \right]
\label{tg}
\end{equation}
(Eq \ref{tg} is a more accurate result than Eq \ref{ttp} for the Tail) 
\begin{equation}
 ({\rm Post-Plateau}) \quad \theta(t) = \thc \left( \frac{t}{\to} \right)^{1/2g}
\label{tpp}
\end{equation}
(for three out of four phases, $\theta \sim t^{1/2}$, while $\theta \sim t^{1/2g}$ increases slower during the post-Plateau phase, 
provided that $g > 1$, which is a requirement for the afterglow Plateau: falling Plateau for $g \in (1,2)$, flat Plateau for $g=2$, 
rising Plateau for $g > 2$)
\begin{equation}
 ({\rm Plateau}) \quad \Gamma(t) = \Gc \left( \frac{t}{\tp} \right)^{-g/2}
\label{Gp}
\end{equation}
and to 
\begin{equation}
 \hh  \Gt = \left\{ \begin{array}{llll} 
 \h   (t/\tg)^{1/2} < 1               & \hh t < \tg                          \\ 
 \h   1                               & \hh \tg \simeq 2\to                      \\ 
 \h   (t/\tg)^{1/2} > 1               & \hh \tg < t                          \\ 
 \h   \Gctc > 1                       & \hh t_p \equiv [(\Gctc)^2+1] \to     \\ 
 \h   (t/t_{pp})^{-\frac{g-1}{2}} > 1 & \hh t_p < t < t_{pp} \quad (g>1)     \\ 
 \h   1                               & \hh t_{pp} \equiv 2\left(\Gctc \right)^{2/(g-1)} t_p   
                                       \\ 
 \h   (t/t_{pp})^{-(g-1)/2g} < 1      & \hh t_{pp} < t                       \\ 
                        \end{array} \right. 
\label{Gt}
\end{equation}
where $\tg$ defined by $\Gt = 1$ represents the end of the GRB phase (observer leaves the cone of maximal radiation beaming),
$\tp$ defined by $\Gt = \Gctc > 1$ being maximal, represents the end of the Tail and beginning of the Plateau (epoch of minimal 
radiation beaming), and $\tpp$ defined by $\Gt = 1$ represents the end of the afterglow Plateau (observer reenters the cone
of maximal radiation beaming), as illustrated in figure 1 of A. Panaitescu (2020). 

 Equation (\ref{Gt}) shows that shows that $\tg=\tp=\tpp/2$ for $\Gctc=1$, which means that outflows with Core parameters $\Gctc \leq 1$
do not have a Tail and a Plateau. Alternatively put, for $\Gctc \leq 1$, which is the maximal value reached by $\Gt$, $\Gt < 1$ at all times,
which means that only the GRB and post-Plateau phases exist (see below). Furthermore, Equation (\ref{Gt}) shows that for $g \lra 1$ 
the Plateau of $\Gt > 1$ extends over an infinite time. Thus, the {\bf fundamental Core parameter $\Gctc$ and Envelope index $g$ can account 
for} GRBs/afterglows with \\
$(i)$ 4 phases (GRB, Tail, Plateau, post-Plateau) if $\Gctc > 1$ and $g > 1$,\\
$(ii)$ 3 phases (GRB, Tail, Plateau) if $\Gctc > 1$ and $g < 1$, and\\
$(iii)$ 2 phases (GRB, post-Plateau) if $\Gctc < 1$,\\
which is almost {\bf the entire diversity displayed by Swift/XRT afterglows}.

 Modeling the burst/prompt and afterglow/delayed emissions with the $LAE$ model determines the temporal milestones $\tp$ and $\tpp$
of those light-curves, which, according to Equation (\ref{Gt}), depend only the combination $\Gctc$ of Core parameters and not on their 
individual values $\Gc$ and $\thc$. Consequently, modeling of the GRB/afterglow light-curves cannot determine both Core parameters,
but only a composite Core parameter $\Gctc$. 
 
 From Equation (\ref{Gt}), one can calculate the Doppler factor
\begin{displaymath}
  D \equiv \frac{1}{ \Gamma (1-(v/c)\cos \theta} \simeq \frac{2 \Gamma}{1 + (\Gt)^2} 
\end{displaymath}
\begin{equation}
   \simeq \left\{ \begin{array}{lll} 
      2\Gamma             & \theta \ll \Gamma^{-1} & ({\rm GRB,\; post-Plateau}) \\ 
      2/(\Gamma \theta^2) & \theta \gg \Gamma^{-1} & ({\rm Tail,\; Plateau}) 
         \end{array} \right.
\label{doppler}
\end{equation}
in the limit $\Gamma >\gg 1$ and $\theta \ll 1$,
which is essential for the calculation of the received flux $F_\eps$ at photon-energy $\eps$
\begin{equation}
   F_\eps (t) \sim  D^2 I'_{\eps/D} \frac{d\theta^2}{dt}
\label{fnu}
\end{equation}
where primed quantities are in the comoving frame of the expanding outflow,
the factor $d\theta^2/dt$ accounts for the area $dA \sim \theta d\theta$ whose emission arrives in a time-interval $dt$ and
the exponent of the Doppler factor for the specific intensity $F_\eps$ is 2 (not 3) because there is no "time-contraction" by 
a factor $D$ when going from comoving-frame to observer-frame, owing to the assumption of an instantaneous emission, with the spread 
$dt$ in photon arrival-time arising only from the angular extent $d\theta$ of an infinitesimal ring $[\theta(t),\theta+d\theta(dt)]$ 
on the radiating surface.

\vspace*{1mm}
\subsection{\bf Flux Calculation}

 The synchrotron and inverse-Compton continua from electrons injected with a power-law distribution with energy and cooling in a constant 
magnetic field are a sequence of three power-law segments. 
For a {\bf a power-law continuum} 
\begin{equation}
   I'_{\eps'} \sim \Ip' \left( \frac{\eps'}{\Ep'} \right)^{-\beta}
\end{equation}
with break-energy $\Ep'$ and break-intensity $\Ip'$, Equation (\ref{fnu}) becomes
\begin{equation}
   F_\eps (t) \sim \left[ D^2 \Ip' \left( \frac{\eps/D}{\Ep'} \right)^{-\beta} \right]_{\theta(t)} \frac{d\theta^2}{dt} 
\label{Fnu}
\end{equation}

 The emission from a patch moving at Lorentz factor $\Gamma (\theta)$ and angle $\theta(t)$ is beamed exactly toward the observer only 
if it was emitted at a comoving angle (relative to the local direction of flow)
\begin{equation}
 \hh  \sin \theta' = D \sin \theta = \frac{2 \Gt}{1 + (\Gt)^2} 
\label{thprime}
\end{equation}
\begin{displaymath}
     \simeq  \left\{ \begin{array}{lll} 
     \hh       2 \Gt     & \hh \theta \ll \Gamma^{-1} & ({\rm GRB,\; post-Plateau}) \\ 
     \hh       2/(\Gt)   & \hh \theta \gg \Gamma^{-1} & ({\rm Tail,\; Plateau}) 
         \end{array} \right.
\end{displaymath}
and in the angle/plane defined by the local direction of flow and the outflow symmetry axis (which, for an on-axis observer, 
is the observer's central line-of-sight toward the center of relativistic expansion). 
The comoving-frame angle $\theta'$ is shown in Figure 1.
 Taking into account that, for highly relativistic electrons of comoving frame Lorentz factor $\gamma$, their emission is beamed 
in the direction of electron motion, it follows that the emission received by the observer from a patch moving at angle $\theta$ 
was produced by electrons moving at angle $\theta'$ relative to the local outward direction of flow, when their gyration around 
the $B$ field direction brought them in the above-described angle/plane.

 Then, the {\bf synchrotron} break-energy and specific intensity at that break-energy satisfy
\begin{equation}
 \Ep' \sim \gamma^2(\theta') B(\theta) \sin\alpha' \;, \; \Ip' \sim \left( \frac{dN}{d\Omega} \right)_{\theta'} B(\theta) \sin\alpha' 
\label{EpIp}
\end{equation}
where $dN/d\Omega$ is the angular density of electrons moving at comoving-frame angle $\theta'$ relative to the local direction of flow,
$B$ is the magnetic field strength in the patch moving at angle $\theta$ relative to the observer and
\begin{equation}
 \cos\alpha' = \cos\theta' \cos\psi' + \sin\theta' \sin\psi' \cos\phi'
\label{aprim}
\end{equation}
is the angle between the electron direction of motion and that of the magnetic field, for a $B$ oriented at angle $\psi'$ relative 
to the direction of flow and an azymuthal angle $\phi'$ for its projection on a plane perpendicular to the flow direction
(prime on $\phi$ is unnecessary because comoving and lab-frame azymuthal angles are identical).

 Consequently, the comoving-frame of the synchrotron $LAE$ is given by 
\begin{equation}
  I'_{\eps'} \eps'^\beta \sim \left( \frac{dN}{d\Omega} \gamma^{2\beta} \right)_{\theta'} [B(\theta)]^{\beta+1}
         \left[ \sin \alpha' (\theta';\psi',\phi) \right]^{\beta+1} 
\label{iprime}
\end{equation}

 For a fixed orientation of the magnetic field, the angle $\phi$ varies from zero to $\pi$ on the equal photon arrival-time 
infinitesimal ring of opening $\theta$. Integration of Equation (\ref{iprime}) over $\phi$ and substitution in Equation (\ref{Fnu}) 
leads to 
\begin{equation}
  F_\eps \eps^\beta \sim \left[ D^{\beta+2} \frac{d\theta^2}{dt} \right]_\theta
                         \left[ \left( \frac{dN}{d\Omega} \gamma^{2\beta} \right)_{\theta'} \left(B^{\beta+1}\right)_\theta \right] 
                         {\cal I}_\beta (\theta',\psi')
\label{Feps}
\end{equation}
\begin{equation}
 {\cal I}_\beta = \int_0^\pi \left[ \sin\alpha'(\phi) \right]^{\beta+1} d\phi = \int_0^\pi (1-\cos^2\alpha')^{(\beta+1)/2} d\phi
\label{Ibeta}
\end{equation}
with $\cos\alpha'$ from Equation (\ref{aprim}) leading to
\begin{equation}
 {\cal I}_\beta = \int_0^\pi \left( a_o + a_1\cos\phi + a_2\cos^2\phi \right)^{(\beta+1)/2} d\phi
\label{a012}
\end{equation}
with $a_0,a_1,a_2$ some coefficients that depend on the angles $\theta'$ and $\psi'$:
\begin{equation}
   a_0 = 1-\cos^2\theta'\cos^2\psi' 
\label{a0}
\end{equation}
\begin{equation}
   a_1 = - \onehalf\sin(2\theta')\sin(2\psi') 
\label{a1}
\end{equation}
\begin{equation}
   a_2 = -\sin^2\theta'\sin^2\psi'
\label{a2}
\end{equation}

 Equation (\ref{Feps}) represents a {\sl basic result} for the calculation of the $LAE$ and illustrates the {\sl factors that determine it}:

$(i)$ The factor in the first square bracket depends on the {\bf angular distribution of the Lorentz factor} $\Gamma$ 
on the instantaneous-emission surface, contains the relativistic boost of that emission, has a {\sl strong} effect on the {\sl received} flux
and, by itself, can account easily for the {\sl four emission phases} (G. Oganesyan et al 2020) and for the {\sl diversity} of GRB/afterglow 
light-curves (A. Panaitescu 2020). 

$(ii)$ The factor in the second square bracket represents the angular structure of three microphysics parameters has a {\sl strong} effect 
on the {\sl comoving-frame emitted} flux and, through departures from axial symmetry, could/should account for short-lived flux variations 
(flares -- R. Duque et al 2022).
These three micro-parameters set four comoving-frame "observables": \\
 $(a)$ the "injection" break-energy $\Ei' \sim \gamma_i^2 B$ at which radiate newly accelerated, uncooled electrons $\gamma_i$, \\
 $(b)$ the "cooling" break-energy $\Ec' \sim \gamma_c^2 B$ at which radiate electrons whose cooling timescale $t'_c \sim (\gamma_c B^2)^{-1}$
       equals the system's age $t'$ (duration of electron injection), \\
 $(c)$ the emission intensities at these two break energies $\Ip' \sim (dN/d\Omega) B$. \\
 Optical and X-ray afterglow observations can determine both the location of and intensity at only one break that is straddled by these domains;
otherwise, optical data can constrain only a combination of location and intensity for a break located below optical.
Consequently, the available data cannot determine all three microphysical parameters but only the characteristics of a single break,
which could be either the injection-break $\Ei'$ or the cooling-break $\Ec'$, with Equation (Eq \ref{EpIp}) applying to the lowest of these 
two breaks, i.e. to the peak of the synchtrotron spectrum.
Thus, we will parameterize (below) the {\bf angular distribution of} that (lowest) {\bf break-energy $\Ep'$ and intensity $\Ip'$}, with this 
detour to a specific radiation mechanism (synchrotron) and its associated microphysical quantities being necessary to extract the dependence 
of the received flux (Eq \ref{Feps}) on the orientation of the magnetic field $\psi'$.

$(iii)$ The last factor (integral ${\cal I}_\beta$) originates in the {\bf anisotropy of the magnetic field} and has a {\sl modest/moderate} 
effect on the {\sl emitted} flux.  

 Each of these factors depend on observer time $t$ through their dependence on the angular location $\theta(t)$ relative to the observer's
line-of-sight toward the center and on the corresponding comoving-frame angle $\theta'(\theta)$ that gets "beamed" toward the observer 
(Eq \ref{thprime}). 
By subsituting these dependences, one can calculate $LAE$ light-curves $F_\eps (t)$ and determine their {\sl temporal decay indices}
as logarithmic derivatives.  

 The integral in Equation (\ref{Feps}) can be calculated analytically only if $\beta$ is an integer, using the usual substitution 
$\tan (\phi/2) = y$ that transforms a combination of trigonometric functions into an algebraic function. 
That substitution is the only path forward if $\beta$ is an even integer, but the integral can be calculated directly, without any substitution, 
if $\beta$ is an odd integer. 

\vspace*{1mm}
\subsubsection{Spectral slope $\beta=1$}

 During the prompt phase, the XRT effective spectral slope (at 0.3-10 keV) softens to $\bx \in (1.5,2.0)$ while being constant $\bx \siml 1.0$ 
during the afterglow phase. Thus, the two integer values of the spectral slope worth of consideration are $\beta=2$ for the prompt emission and 
$\beta=1$ for the afterglow. Unfortunately, the former case leads to an obfuscating algebraic integrand and will not be pursued here, where we 
proceed with the latter case, which is representative for the typical afterglow X-ray spectral slope. For that case, one obtains
\begin{equation}
 \eps F_\eps (t) \sim \frac{d\theta^2}{dt} (D^3 B^2)_\theta  \left(\frac{dN}{d\Omega}\gamma^2 \right)_{\theta'} 
       X\left[\theta'\left(\theta(t)\right),\psi'\right] 
\label{FXpsi}
\end{equation}
with 
\begin{equation}
 X(\theta',\psi') = \sin^2\theta' \cos^2\psi' + \left( 1-\onehalf \sin^2\theta' \right) \sin^2\psi' 
\label{Xcorr}
\end{equation}

 This {\bf $B$-anisotropy correction factor} $X(t,\psi')$ is shown in Figure 1 for a few orientation angles $\psi'$, is a function of time,
and represents, obviously, the correction introduced by a uni-directional magnetic field on the received flux (calculated previously by 
G. Oganesyan et al 2020 and A. Panaitescu 2020, 2025) for an isotropic $B$, as shown by integrating $X(t,\psi')$ over an omni-directional $B$ 
\footnote{An {\bf isotropic magnetic field} means 
   a field that changes direction randomly on the infinitesimally-thin ring of equal photon-arrival time}:
\begin{equation}
 \int_0^\pi X(\theta',\psi') \left[ P(\psi') \sim \sin\psi'\right] d\psi' = 4/3
\end{equation}
Note that the dependence on the comoving-frame emission angle $\theta'$ has disappeared, thus there is no received-flux time-dependence associated 
with an isotropic magnetic field.

 By isolating the term that contains $\theta'(t)$: $X(\theta',\psi')=(\cos^2\psi'-\onehalf \sin^2\psi')\sin^2\theta' + \sin^2\psi'$, we can 
identify a "magic" $B$-angle $\Psi' = \arctan \sqrt{2} = 54.7^\deg$ for which the $\theta'$ dependence vanishes, which means a correction factor 
$X(\Psi')=2/3$ that is time-independent, just as for an isotropic $B$. 

 A minor note is the $90^\deg$-symmetry of the factor $X(\psi')$, which is due to the $90^\deg$-symmetry of coefficients $a_0$ and $a_2$ 
of Equations (\ref{a0}) and (\ref{a2}), which "survive" integration over the angle $\phi$, while the $90^\deg$-anti-symmetric coefficient $a_1$ 
disappears after the $\phi$ integration. That $X(\psi') = X(\pi-\psi')$ implies that we can safely restrict attention to $\psi' \in [0,\pi/2]$.

 The first term in the right-hand-side of Equation (\ref{Xcorr}) represents the effect of a parallel $B$ ($\psi'=0$) on the $LAE$ flux and 
its coefficient $X_{pAr}=\sin^2\theta'$ varies between 0 and 1.
The second term in the rhs contains the effect of a perpendicular $B$ ($\psi'=90^\deg$) and its coefficient $X_{pEr}=1-\onehalf \sin^2\theta'$ 
varies between 0.5 and 1.
These indicate that {\sl a parallel $B$ has a stronger effect} on the $LAE$ {\sl than a perpendicular $B$}, as illustrated in Figure 1,
but the difference easiest to measure between these two $B$ orientations is in the asymptotic power-law flux-decay indices (next) and
not from the long-term flux fluctuations that they induce during the afterglow phase. 

 Using Equations (\ref{doppler}) and (\ref{thprime}), it can be shown that the parallel-$B$ and perpendicular-$B$ correction factors 
are equal for a "critical" angle $\psicr$ satisfying
\begin{equation}
 \tan \psicr (t) = \frac{\sin\theta'}{\sqrt{1-\onehalf \sin^2\theta'}} = \frac{2\Gt}{\sqrt{1+(\Gt)^4}}  
\label{psicr}
\end{equation}
\begin{displaymath}
   \simeq  \left\{ \begin{array}{lll} 
    \hh   2 \Gt     & \theta \ll \Gamma^{-1} & ({\rm GRB,\; post-Plateau}) \\ 
    \hh   2/(\Gt)   & \theta \gg \Gamma^{-1} & ({\rm Tail,\; Plateau}) 
           \end{array} \right.
\end{displaymath}
Comparison with Equation (\ref{thprime}) shows that $\tan \psicr \simeq \sin \theta'$ when $\Gt$ is sufficiently far from unity, 
i.e. far from the GRB end/Tail start at $\tg$ and Plateau end/post-Plateau beginning at $\tpp$.

 Replacing $\Gamma(t)$ and $\theta(t)$ in above equation leads to expressions for $\psicr(t)$. The variable critical angle $\psicr$ may cross 
a fixed magnetic-field orientation $\psi'$ at four epochs, one for each GRB/afterglow phase, that can be calculated for any $\Gt$. 
Each crossing represents a switch between dominant terms in the correction factor $X(t,\psi')$, as illustrated in Figure 1.
That switch happens only if $\tan \psi' < \sqrt{2}$. Furthermore, the switch happens at times of interest only if $\tan \psi' > 2/(\Gctc)$
(for $\Gctc \gg 1$), otherwise the $B$-parallel correction-term is always dominant for $\psi' < \arctan (2/\Gctc)$, while the $B$-perpendicular 
correction-term is always dominant for $\psi' > \arctan \sqrt{2}$.

\vspace*{1mm}
\subsection{\bf Flux-Decay Index: Parallel $B$}

 The integration shown in Equation (\ref{a012}) is not necessary for a parallel magnetic field ($\psi'=0$) because $\alpha' = \theta'$ 
in that case. That means that a spectral slope $\beta = 1$ need not be assumed for Equation (\ref{Ibeta}). 
 Substitution of $\Gamma(t)$ (Eqs \ref{Gpp} and \ref{Gp}) and $\theta(t)$ (Eqs \ref{ttp}, \ref{tg}, and \ref{tpp}) in Equations (\ref{doppler}) 
and then in Equation (\ref{Feps}) leads to
\begin{equation}
  F_\eps (t) \sim t^{-\alpha} \eps^{-\beta} \; 
\end{equation}
with the following {\bf asymptotic indices of the power-law $LAE$ flux-decay} in the {\bf limits} $\Gt \ll 1$ (for $t \ll \tg$ and $\tpp \ll t$)
and $\Gt \gg 1$ (for $\tg \ll t < \tp$ and $\tp < t \ll \tpp$)
\begin{equation}
 \hhh \alpha_{pAr} \h = \h \frac{1}{2} \h \left\{ \begin{array}{lll} 
  \hh -(1+\beta)                   & \hh t \ll \tg        & \hhh (\ag) \\ 
  \hh  5+3\beta                    & \hh \tg \ll t < \tp  & \hhh (\at) \\
  \hh  5+3\beta - g(3+2\beta) - x  & \hh \tp < t \ll \tpp & \hhh (\ap) \\
  \hh  5 + 2\beta - (3+\beta+x)/g  & \hh \tpp \ll t       & \hhh (\app) 
  \end{array} \right.
\label{Bpar}
\end{equation}
where the break-energy $\Ep'$ and break-intensity $\Ip'$ were assumed to be angle-independent in the uniform Core and have a power-law
angular structure in the Envelope (just like the outflow Lorentz factor)
\begin{equation}
 \Ip'(\theta) \sim \theta^i , \; \Ep'(\theta) \sim \theta^e \lra I'_{\eps'}(\theta) \sim \theta^x \;,\; x \equiv i + \beta e
\label{inu}
\end{equation}
Note that a parallel $B$ yields a rising GRB light-curve, owing to a constant Doppler factor $\calD = 2\Gamma_c$ and a rising 
flux-correction factor $X_{pAr} \sim (\sin \theta')^{\beta+1} \sim (D \sin \theta)^{\beta+1} \simeq \theta^{\beta+1} \sim t^{(\beta+1)/2}$. 
However, the rise of the Doppler factor is short-lived and the GRB flux begins to fall-off soon, displaying a peak before the end 
of the GRB phase.

\vspace*{1mm}
\subsection{\bf Flux-Decay Index: Perpendicular $B$}

 A similar substitution of $\Gamma(t)$ and $\theta(t)$ in Equation (\ref{Feps}) allows the derivation of the power-law flux-decay indices
for a perpendicular $B$ ($\psi'=90^\deg$) and for $\beta=1$ 
\begin{equation}
  \alpha_{pEr} = \left\{ \begin{array}{lll} 
  \h  0                  &  t \ll \tg        \\ 
  \h  3                  &  \tg \ll t < \tp  \\
  \h  3 - 3g/2 - x/2     &  \tp < t \ll \tpp \\
  \h  2.5 - 1/g - x/2g   &  \tpp \ll t
  \end{array} \right.
\end{equation}
 Following how the spectral slope $\beta$ enters in the calculation of the $LAE$ flux (and its asymptotic decay indices) for a parallel $B$,
one can "creatively" generalize the above decay indices for any spectral slope $\beta$
\begin{equation}
 \hhh \h \alpha_{pEr} \h \simeq \h \left\{ \begin{array}{lll} 
  \hh  0                             & \hh t \ll \tg        & \hhh (\ag) \\ 
  \hh  2+\beta                       & \hh \tg \ll t < \tp  & \hhh (\at) \\
  \hh  2+\beta - g(1+\beta/2) - x/2  & \hh \tp < t \ll \tpp & \hhh (\ap) \\
  \hh  2 + \beta/2 - 1/g - x/2g      & \hh \tpp \ll t       & \hhh (\app)
  \end{array} \right.
\label{Bper}
\end{equation}
Note that a parpendicular $B$ yields a flat GRB light-curve, due to the constant Doppler factor and to the weak variation with time 
of the flux-correction factor $X_{pEr} = 1-\onehalf \sin^2 \theta'$.

 The asymptotic Tail flux-decay index $\at = 2+\beta$ was derived by P. Kumar \& A. Panaitescu (2000) under the tacit assumption of an 
isotropic magnetic field, for which a perpendicular $B$ is a good proxy (see below). But note that a parallel $B$ yields a Tail-flux
decay-index (Eq \ref{Bpar}) "faster" by $(\beta+1)/2 \siml 1.5$ for $\beta \siml 2$ at the end of the GRB Tail.

\vspace*{1mm}
\subsection{\bf Effect of $B$ orientation}

 Figure 2 illustrates the effect that the $B$ orientation angle $\psi'$ has on the $LAE$ light-curve (left panel) and on the asymptotic 
flux-decay indices (right panel). The important conclusions to be drawn are that a parallel $B$ decreases the flux at the beginnings and
ends of the prompt and delayed phases while a perpendicular $B$ decreases the flux at the logarithmic middle of each phase, with the
net effect that {\bf a parallel $B$ increases the curvature of the light-curves}, with "curvature" defined by the difference between
asymptotic decay indices for each phase.

 The second conclusion to be drawn from Figure 2 is that the asymptotic decay indices for larger angles $\psi'$ are similar, which is due 
to the weak time-variation of the coefficient $X_{pEr}=1-\onehalf \sin^2\theta'$ of the perpendicular-$B$ correction-factor $X(\psi')$.
 This indicates that, from the point of view of asymptotic indices, the {\sl dichotomy introduced by the orientation of the magnetic field}
is one {\sl between a parallel field and perpendicular/isotropic one}, with the flux-decay indices for a perpendicular $B$ being representative 
for all large orientation angles $\psi'$, including even the magic angle $\Psi'=55^\deg$, for which $X(t)$=constant, which is also a good 
stand-in for an isotropic $B$.    

 The above decay indices given in Equation (\ref{Bper}) for a perpendicular $B$ are exactly those derived by A. Panaitescu (2020) 
under the unrecognized assumption of an isotropic magnetic field.

\vspace*{2mm}
\section{\bf Important Details}

 The following subsections present some important aspects about the $LAE$ model and its application to the optical and X-ray emissions
of the seven GRBs shown in Figures A3-A9.

\vspace*{1mm}
\subsection{\bf Model Parameters}

  The $LAE$ model has \\
$(i)$ {\sl two fundamental dynamical parameters} for the uniform Core's $\Gctc$ and for the Envelope's Lorentz-factor angular distribution 
$\Gamma \sim \theta^{-g}$,  \\
$(ii)$ {\sl three fundamental emission parameters} for the Envelope's characteristics angular distributions: comoving-frame break-energy 
$\Ep' \sim \theta^e$ and intensity $\Ip' \sim \theta^i$ at $\Ep'$, and orientation $\psi'$ of the magnetic field, \\
$(iii)$ {\sl six secondary emission parameters} for the Core and Envelope normalization of the spectral break-energy and flux at that break
($\Ep^{(c)}, \Fp^{(c)}; \Ep^{(e)},\Fp^{(e)}$) and high-energy slope ($\bH^{(c)},\bH^{(e)}$ above it, and \\
$(iv)$ {\sl two secondary temporal parameters} for the epoch $\tej$ when the ejecta were released (which is never the Swift/BAT trigger) 
and when the instantaneous emission is produced ($\to>\tej$).

\vspace*{1mm}
\subsubsection{Emission epoch $\to$} 

 Given that, in this treatment of the $LAE$ model, we assume an instantaneous emission release at $\to$ and ignore any prior emission, 
{\sl fixing} $\to$ at the light-curve peak is well-motivated because an earlier choice would yield poorer fits, while a later choice would 
discard measurements and thus reduce the ability of the XRT data to constrain the $B$ field orientation. For most GRBs considered here,
the XRT has caught the peak of the 0.3-10 keV prompt emission. If that epoch was missed by the XRT, then the BAT light-curve may indicate
the peak-epoch $\to$. In that case, various fixed values for $\to$ are considered, to assess its effect on the determination of the other
model parameters. 

 The $\to$ could be a free model parameter and could be determined from requiring that the Tail power-law flux decay at 0.3 keV and 10 keV 
is the expected power-law. E. Liang et al (2006) have done so to the 0.3-10 keV light-curves of several GRBs, using the Tail flux-decay 
$t^{-(\bx+2)}$ expected for a perpendicular field. However, allowing a $\to$ other the GRB pulse peak introduces an unwanted bias of 
favoring a parallel $B$ if earlier measurements are included, or of favoring a perpendicular $B$ when later measurements are excluded.

\vspace*{1mm}
\subsubsection{Ejection epoch $\tej$}

 In all preceeding equations, time is measured from the unkown epoch $\tej$ when the outflow producing the last major GRB pulse was launched.
In fitting light-curves, we use the GRB end epoch $\tg$ instead of the ejection epoch $\tej$ because the former allows an easy visual estimation.
First, with $\tg$ defined by $\Gt=1$ (the observer exits the cone of maximal relativistic beaming at the end of the GRB phase and beginning 
of the Tail), Equation (\ref{Gt}) shows that $\tg-\tej \simeq 2(\to-\tej)$, hence $\tej \simeq \to - (\tg-\to)$.
Further, Equations (\ref{Bpar}) and (\ref{Bper}) for the Tail flux decay $\at$, suggest that, by the end of the GRB phase, the $LAE$ has
dimmed by about half of $(\tg-\tej)/(\to-\tej)^{(\beta+1)|(2.5+1.5\beta)}$, i.e. by a factor 5-20 from the peak flux (for $\beta=1.2$).
For most best-fits found here, the GRB end epoch $\tg$ is when the GRB flux has dropped by a factor 10 from its initial value,
thus the ease in estimating $\tg$ visually. Because these are just estimates, the epoch $\tg$ is a {\sl free} model parameter.

\vspace*{1mm}
\subsubsection{Spectral slopes}

 The low-energy spectral slope below the break at $\Ep$ of the {\sl Core/prompt emission} is {\sl fixed} at $\bL = -1/3$ (uncooled electrons) 
because the hardest possible synchrotron emission is, sometimes, directly required by the measured XRT effective slope $\bx$ at 0.3-10 keV. 
In the remaining cases, a softer low-energy slope $\bL = 1/2$ (cooled electrons) is compatible with XRT measurements for $\bx$ but always 
yields a too-soft effective slope because the best-fit break-energy $\Ep$ is too low (in the XRT window) and/or evolves too slowly (by fiat).
Thus, while a quick trial of $\bL = 1/2$ never hurts, it also never helps, most likely owing to the limitations associated with a uniform 
Core assumption (see below).

 The low-energy slope of the {\sl Envelope/delayed emission} is constrained {\sl only} by the optical data and not by the X-ray measurements,
which are at a photon energy above $\Ep$ after the prompt phase. Both expected, {\sl fixed} values $\bL = -1/3,1/2$ are allowed but, by chosing 
for modeling only the brightest optical afterglows (a very natural selection bias), one favors the softer low-energy slope $\bL = 1/2$ 
of cooled electrons. 

 The high-energy slopes $\bH$ above the break-energy $\Ep$ (for the Core and Envelope) are {\sl free} (and constant) model parameters 
and are over-determined observationally by 
$(i)$ the flux decays measured for the Tail ($\at$) and the afterglow ($\ap,\app$), as shown by Equations (\ref{Bpar} and (\ref{Bper}),
and by $(ii)$ the XRT effective slope $\bx$.

\vspace*{1mm}
\subsubsection{Break-energies and break-intensities}

 The normalizations of the break-energies $\Ep$ and corresponding intensities $\Ip$ for both the Core/prompt and Envelope/delayed emissions 
are {\sl free} model parameters and are determined by the relative 0.3 and 10 keV fluxes during the prompt phase (when $\Ep$ is in the XRT 
band) and by the relative optical and X-ray fluxes during the afterglow phase (when $\Ep$ is between optical and X-ray).

 Given the discontinuity of the high-energy spectral slope $\bL$ at the end-of-prompt/beginning-of-afterglow (epoch $\tp$) in the $LAE$ model 
(the XRT effective spectral slope may display a gradual hardening of the effective slope $\bx$ before $\tp$, but that feature cannot be captured 
in a model with a uniform Core, see below), there is little point in enforcing a continuous transition of the spectral characteristics
$\Ep$ and $\Ip$ at the Core-Envelope boundary. 

 Furthermore, such a continuous transition, which would imply that the break-energy $\Ep$ is just below 0.1 keV after the prompt phase, 
can easily be incompatible with the high brightness of the early optical afterglow, which requires a break-energy $\Ep$ that is closer 
to the optical than to the X-ray. Our selection of optical afterglows with a good temporal coverage introduces a selection bias toward 
the brightest optical afterglows, which are 10-100 times brighter than in the X-ray. For a high-energy slope $\bL \siml 1.0$, that high
optical-to-X-ray flux ratio (which should be less than $(0.3\, keV/3\, eV)^{\bH}$) requires a break-energy $\Ep$ that cannot much above optical. 

 While the best-fit value for the break-energies $\Ep$ and intensities $\Ip$ can be determined from fitting optical and X-ray light-curves, 
those values cannot be converted to the more basic outflow quantities ($dN/d\Omega, \gamma_i, B$) of Equation (\ref{Feps}) because
they set two observational constraints for three basic parameters. A third observational constraint could be the determination of the
electron cooling-break $\Ec$ if $\Ep$ is the electron injection-break $\Ei$ (or of the injection-break if $\Ep$ is the cooling-break),
which would require observations in a third photon-energy domain. Early radio observations (within the first day) are not available and,
even if they were, they would be marred by a large interstellar scintillation, and not attributable to the $LAE$ because the GRB ejecta
should be optically thick in the radio during the first day. Fermi observations of the early afterglow above 100 MeV may be available
ocassionally, but it is unlikely that the cooling-break energy $\Ec$ is above 10 keV.

\vspace*{1mm}
\subsection{\bf Basic Parameters -- Observational Constraints}

 Equations (\ref{Gt}), (\ref{Bpar}), and (\ref{Bper}) show that the X-ray emission (prompt and delayed) provides {\bf five} observational constraints 
($\tp/\tg,\tpp/\tp;\Dagrb=\at-\ag,\ap^{(x)},\app^{(x)}$) for {\bf three} $LAE$ model basic parameters ($\Gctc;\psi';g$) and a composite of {\bf two}
parameters ($i+e\bx$) that is degenerate in the fundamental parameters ($i,e$).
 Thus, X-ray measurements provide a number of observational constraints larger than that of free model parameters, which raises the prospects 
of testing the $LAE$ model (with all its current assumptions/limitations) and of determining (i.e. constraining to small ranges) three of its five
fundamental parameters. In practice, the former is thwarted by large systematic fluctuations during both the prompt and delayed phases,
pointing to possible angular fluctuations of all quantities ($\Gamma,dN/d\Omega, \gamma_i, B$) that determine the received flux.

 The afterglow optical emission provides up to {\bf three} observational constraints ($\tpp/\tp;\ap^{(o)},\app^{(o)}$), with the temporal milestones 
useful at confirming if the optical emission  originates from the same mechanism as the X-ray (in which case, light-curves should be well-coupled,
with achromatic breaks occurring at the same Plateau end epoch $\tpp$). 

 Thus, optical and X-ray coupled light-curves provide {\bf seven} independent observational constraints ($\tp/\tg,\tpp/\tp;\Dagrb,\ap^{(x)},\app^{(x)};
\ap^{(o)},\app^{(o)}$) for {\bf five} $LAE$ model parameters ($\Gctc;\psi';g;i,e$).
Optical measurements have lower systematic fluctuations than X-ray observations, but the testing ability of $LAE$ fits afforded by having more 
observational constraints than relevant model parameters remains impaired by the X-ray fluctuations.

 Instead, the ability of the $LAE$ model to account for the X-ray emission of GRBs and their afterglows should be evaluated by the ease at which
this model unifies the prompt and delayed emissions using a single "mechanism" (G. Oganesyan et al 2020) and by the ease at which it can account 
for the diversity of GRB/afterglows using only one Core parameter ($\Gctc$) and one Envelope parameter (Lorentz factor index $g$) (A. Panaitescu 2020).
 Furthermore, the $LAE$ model may also account for the optical emission of those optical light-curves that are coupled to the X-ray, if allowance
for discontinuities in the emission characteristics ($\bH;\Ep',\Ip'$) is made. 

 Evidently, having more observational constraints than model parameters promises a better chance of determining the model parameters,
fits to the optical and X-ray data will be used only for this purpose.
To identify how much an observational component (prompt vs. delayed, optical vs. X-ray) constrains the fundamental $LAE$ model parameters, 
prompt/X-ray or afterglow/optical measurements will be alternatively or progressively excluded from the fit. 

\vspace*{1mm}
\subsection{\bf Limitations of a Uniform Core}
\label{limits}

 Assuming a uniform Core, where all parameters are angle-independent, eases analytical derivations but is also necessary for calculating fluxes 
with a minimal number of model parameters while avoiding an on-axis diverging (power-law) profile (a Gaussian or exponential angular structure 
would naturally do that but are not sufficiently different from a uniform profile to alleviate the issues below). 
       
 One issue is that, for a uniform angular distribution, the prompt-emission softening is due only to the decreasing Doppler factor: 
$\Ep = \calD \Ep' \sim \Ep'/t$. 
 A break-energy $\Ep$ decreasing slower than observed slightly increases the flux-decay at an energy below $\Ep$ if $\bL = -1/3$ 
(and mitigates that decay if $\bL=1/2$), but reduces substantially the flux-decay above $\Ep$ (because $\bH < -1$).
Put together, it follows that a slower decrease of $\Ep$ leads to a smaller light-curve curvature $\Dagrb = \at-\ag$ (the difference 
between the asymptotic power-law decay indices at beginning of the GRB and at the end of the Tail), biasing the $LAE$ model toward 
a parallel magnetic field, to acquire the light-curve extra curvature required by observations.
       
 Another issue pertains to GRB Tails that display a spectral {\bf hardening} prior to the Plateau (clearly shown by GRB 060729, 060714, 060526, 130427), 
which a uniform Core, where the break-energy $\Ep'$ and high-energy slope $\bH$ are angle-independent, cannot accommodate. That hardening requires 
either an energy $\Ep'$ that increases with angle $\theta$ or a slope $\bH$ that decreases with $\theta$. 

 The former possibility seems untenable because 
the hardening seen at the end of the Tail is followed by a constant hardness throughout the afterglow, which requires that $\Ep \calD \Ep'$
remains constant or stays above 10 keV; those circumstances may be possible during the Plateau, when $\calD$ is constant or very slowly 
evolving (which is the $LAE$ explanation for Plateaus), but not likely during the post-Plateau, when $\calD \sim \Gamma \sim t^{-1/2}$. 
Furthermore, the effective spectral slope $\bx \siml 1$ usually measured by XRT during the afterglow is too soft even for the softest
possible low-energy slope $\bL = 1/2$, which indicates that the break-energy $\Ep$ is not above 10 keV during the afterglow.

 Instead, it is more likely that the pre-Plateau hardening is due to a decreasing high-energy slope $\bH$ that compensates partially
for a decreasing break-energy $\Ep$ and yields an overall spectral hardening in the XRT band. Such a feature (a slope $\bH$ decreasing 
during the Tail) is not included in the $LAE$ model because it is ad-hoc and because that hardening starts at an arbitrary epoch 
(introducing thus a new model parameter), although it seems to be required often by XRT observations.

 A pre-Plateau spectral hardening originating in a decreasing slope $\bH$ at the Tail end mitigates the decay of the 10 keV flux but 
leaves unchanged the 0.3 keV light-curve. That differential effect decouples the two X-ray light-curves beyond what the $LAE$ model 
can accommodate. Furthermore, the pre-Plateau spectral hardening reduces the curvature of the 10 keV flux, 
which biases a model with a uniform slope $\bH$ toward a perpendicular magnetic field, to reduce the curvature of the model light-curves.

 In summary, two assumptions of the $LAE$ model -- a uniform Core and a spectrum of constant high-energy spectral slope --
bring in two caveats for the determination of the magnetic field orientation from the curvature of the prompt phase light-curves: \\
 $(1)$ for a {\sl Tail displaying a $\Ep$ decrease} faster than the $LAE$ model expectation $\Ep \sim t^{-1}$, the $LAE$ model 
will be {\sl biased toward a parallel magnetic field} in the Core. More specifically, when the measured energy $\Ep$ falls-off faster than 
$t^{-1}$, determinations of a Core perpendicular $B$ from the curvature of the prompt emission light-curves are safe, while determinations 
of a Core parallel $B$ could be unreliable. \\
 $(2)$ for a {\sl Tail displaying a spectral hardening} prior to the Plateau, the $LAE$ will be {\sl half-biased toward a perpendicular magnetic 
field} in the Core. In other words, when the Tail end displays a spectral hardening, determinations of a Core parallel $B$ from the curvature
of the prompt emission light-curves are safe, while determinations of a Core perpendicular $B$ can be unreliable.

 Furthermore, either type of softening (a measured break-energy $\Ep$ falling-off faster than $t^{-1}$ during the Tail, or a measured effective 
spectral slope $\bx$ hardening before the Plateau) leads to poorer fits with a uniform Core $LAE$ model. Still, the determination of the Core 
parameter $\Gctc$, which is based on the Tail timing, remains robust.

\vspace*{1mm}
\subsection{\bf A Sample Selection Effect} 
\label{offaxis}

 To determine all fundamental parameters of the $LAE$ model ($\Gctc;g;\psi';e,i$) requires X-ray light-curves with all four phases
(GRB, Tail, Plateau, post-Plateau) well-monitored, so that all observational constraints ($\tp/\tg,\tpp/\tp;\Dagrb,\ap^{(x)},\app^{(x)}$) 
are used. That selection biases fitting X-ray light-curves toward observer locations close to the outflow symmetry axis (as described below),
which is overall a good thing because an off-axis observer location $\thobs$ precludes analytical results in a explicit simple form 
like those in Equations (\ref{Bpar}) and (\ref{Bper}). 

 A minor set-back arising from adding another free parameter ($\thobs$) could be that is slows down the multi-parameter 
($\to,\tg; \bL^{(c)},\Ep^{(c)}, \Fp^{(c)}; \bL^{(e)},\Ep^{(e)},\Fp^{(e)}; g; \psi'; e,i$) search for the best-fit, where $^{(c)}$ stands for Core 
and $^{(e)}$ for Envelope.
 
 A good reason for including the offset angle $\thobs$ is that it may yield wider ranges for the best-fit fundamental parameters 
although a narrow parameter range is not an issue for the best-fits below.
The best reason for that inclusion is that it may lead to better fits, but that is not the case with the best-fits below, whose poorness
arises from the chromatic features and variability of the X-ray light-curves, neither of which can be explained by the observer offset.

 A practical reason against including the {\sl observer offset angle} is that it {\sl has no physical relevance}, being just a dummy parameter, 
albeit an effective one. All that one can do with a set of best-fit $\thobs$ is to test the assumption of a uniform Core by checking that 
its distribution follows the expected $P(\thobs) \sim \sin \thobs$. 

 The higher brightness of the Core emission implies a bias toward detecting GRBs for which the observer is located withing the Core opening
($\thobs < \thc$), while off-Core observer locations ($\thobs > \thc$) could lead to orphan afterglows, whose GRB emission was not detected.
For the former case, if the Core is uniform, then the flux received from infinitesimal rings of ever-increasing opening around the direction 
toward the observer has no knowledge of the offset angle $\thobs$ and the GRB flux decay at early times has the same asymptotic decay index 
$\ag$ as given in Equations (\ref{Bpar}) and (\ref{Bper}). Departures from that flux-decay start when the observer begins to "see" the 
Core-Envelope boundary, i.e. when the equal photon-arrival time ring reaches an opening $\theta(t) = \thc-\thobs$. For $\thc-\thobs > \Gc^{-1}$ 
that will happen after $\Gt > 1$, i.e. after the GRB phase and during the Tail. After $\theta(t) = \thc-\thobs$  and until $\theta(t) = \thc+\thobs$,
the flux received is a mixture of the brighter Core emission that would yield the fast-decaying Tail for an on-axis ($\thobs=0$) observer 
and the dimmer Envelope emission that would yield the flat Plateau for the same privileged observer. 

 Thus, the effect of an off-axis, but in-Core, observer location ($\thobs < \thc$) is to "move" the brighter Tail-end emission to later times
and the dimmer Plateau-beginning emission to earlier times, ironing out more and more of the Tail-Plateau transition for an increasing
offset $\thobs$, Consequently, $LAE$ light-curves for $\thobs \simeq \thc$ should lack the entire Tail and most of the Plateau, as illustrated 
in figure 7 of S. Ascenzi et al (2020). 

 The above effect of the observer offset suggests another way of explaining some of the diversity of X-ray afterglows: some afterglows 
lack a Tail and a Plateau not because the Core parameter $\Gctc \leq 1$ (Eq \ref{Gt}), but because the observer is located closer 
to the Core-Envelope boundary than to the outflow symmetry axis.

 That effect also illustrates why selecting for modeling only GRB/afterglows with well-developed Tails and Plateaus sets a bias toward observer 
locations close to the outflow axis.

\vspace*{3mm}
\section{\bf Conclusions}

\vspace*{1mm}
\subsection{\bf Afterglow diversity}

 Developing an analytical model for the $LAE$ and applying it to fit the well-coupled (similar flux decays, synchronous temporal breaks) optical 
and X-ray light-curves of several GRBs and their afterglows was not done to test that model because it has enough flexibility to accommodate any 
"reasonable" afterglow light-curve. 
 That flexibility arises from the Core parameter $\Gctc$ (Core angular size $\thc$ times Core Lorentz factor $\Gc$) and from the Envelope dynamical 
parameter $g$ for the power-law angular distribution of the Lorentz factor $\Gamma(\theta) \sim \theta^{-g}$.

 Those parameters are enough to account for most of the diversity displayed by X-ray prompt and delayed light-curves: 
 $(i)$ four phases (GRB, Tail, Plateau, post-Plateau) if $\Gctc > 1$ and $g > 1$, 
 $(ii)$ three phases (GRB, Tail, and Plateau by its $\Gt > 1$ definition, but post-Plateau by its fast flux-decay) if $\Gctc > 1$ and $g < 1$, 
 $(iii)$ two phases (GRB and post-Plateau) if $\Gctc < 1$ (for any $g$). 
It is worth noting that the second type of afterglows could be due to an in-Core observer location that is far from the outflow symmetry axis. 

 To the above, one could add that X-ray fluctuations can be attributed to hot-spots on the emitting surface, i.e. to departures from the assumed 
axial symmetry, and other rare/bizarre light-curve features, such as sudden drops, can be attributed to departures from smooth angular dependences 
of the dynamical and emission-characteristics indices of an axially-symmetric outflow.

\vspace*{1mm}
\subsection{\bf $LAE$ parameters and observational constraints}

 Instead of model testing, fits with the $LAE$ model to well-coupled (but not completely coupled) optical and X-ray light-curves were done 
(1) to illustrate the ability and the limitations of the model's major assumptions (most notable, the assumption of a uniform Core) and
(2) to determine its fundamental parameters ($\Gctc; g; e,i,\psi'$), including the indices $e$ and $i$ for the power-law angular dependence
of the break-energy $\Ep$ and intensity $\Ip$ at that energy, with $\psi'$ being the angle between the magnetic field and the radially out direction
of flow, and with the purpose of identifying if the magnetic field is parallel, perpendicular, or isotropic.

 The emission from a uniform Core is less flexible, having only two parameters ($\Gctc$ and $\psi'_C$); the emission from a power-law Envelope has
more flexibility, having four free parameters ($g; e,i,\psi'_E$), with the $B$ field orientation potentially being diferent from Core to Envelope.  

 If there is no spectral break between optical and X-rays, the corresponding afterglow light-curves are perfectly coupled, displaying the same flux 
power-law decay at all times, and setting only two observational constraints (the asymptotic power-law decay indices $\ap$ at the Plateau beginning 
and $\app$ well after the Plateau end -- Eqs \ref{Bpar} and \ref{Bper}). In this case, the model parameters cannot be determined (i.e. well-constrained). 
 That parameter determination can be achieved only if at least one spectral-break is between optical and X-ray, which leads to well-coupled light-curves 
and to four observational constraints ($\ap^{(o,x)}$ and $\app^{(o,x)}$ in the optical and in the X-ray).

\vspace*{1mm}
\subsection{\bf Core dynamical parameter}

 Both prompt and afterglow flux measurements determine the Core parameter $\Gctc$ which sets the temporal milestones $\tg,\tp,\tpp$ of the GRB and afterglow
and the dynamical ranges for the Tail ($\tp/\tg = (\Gctc)^2$) and for the Plateau ($\tpp/\tp = 2(\Gctc)^{2/(g-1)}$). Panels C1 of Figures A3-A9 show that
afterglow optical and X-ray observations alone lead to an uncertain $\Gctc$, which is due to the uncertainty in determining the Envelope index $g$, 
but adding the prompt X-ray data provides a better determination of that Core parameter, restricting the allowed range to $\Delta \Gctc \siml 1$ in about 
half the GRBs fit in this work. 

  On hydrodynamical grounds, the lateral expansion of the outflow Core is expected to add a lateral spread of $1/(\sqrt{3}\Gc$. Thus, any Core of initial 
opening $\thc > 0.4/\Gc$ would have an opening $\thc > 1/\Gc$ at the site when the instantaneous emission is produced. Our selection of GRBs with all four 
phases well-monitored ensures that $\Gctc > 1$. Excluding the larger value of that Core parameter inferred here for GRB 060614, the remaining six Cores 
have an average $\Gctc=3.0$, and that value is compatible with all the allowed intervals identified in Figures A3,A4,A6-A9, i.e. $\Gctc=3$ could be a 
quasi-universal value, with {\sl faster jets having narrower Cores}.

 The simulations of G. Urrutia, F. De Colle \& D. Lopez-Camara (2023) of jets propagating through massive progenitors indicate a correlation between 
the jet-driving mechanism and jet properties, with pressure-dominated jets being less relativistic (lower $\Gc$) and less collimated (higher $\thc$) 
than kinetic jets.  That could imply a $\Gc-\thc$ anticorrelation that would explain why the Core parameter $\Gctc$ may have a universal value.

\vspace*{1mm}
\subsection{\bf Index of Envelope Lorentz-factor distribution}

 Panels C2 of Figures A3-A9 show that a tight determination ($\Delta g < 1$) of the {\bf Envelope} index $g$ for the angular distribution of the Lorentz factor, 
$\Gamma(\theta) \sim \theta^{-g}$, is achieved for about half of the seven GRBs fit here. The substantial uncertainty in the index $g$ for the remaining cases
is due to that the asymptotic afterglow flux-decay indices $\ap$ and $\app$ are also set by the emission characteristics indices $e$ and $i$, which also 
have large uncertainties (below). 

 Unsurprisingly, the average of all seven mean best-fit indices is $g=2.1$, which is the analytically expected value for a flat light-curve Plateau. 
Envelopes with a shallow $g < 1$ structure yield decays that are too fast to be called Plateaus, while Envelopes with a strong $g > 3$ structure yield
rising light-curves. Our choice of afterglows with (flat) Plateaus excludes such extreme Envelope structures.
The allowed indices $g$ span the interval $(1.5,2.5)$ for most GRBs considered here.


 Numerical calculations of jet acceleration by conversion of internal energy into kinetic led P. Duffell \& A. MacFadyen (2013) to a steep power-law 
structure with index $g \simeq 1.5$ under certain initial conditions (their fig 3). 

 The results of hydrodynamical simulations of relativistic jets interacting with the remainder of the stellar progenitor are often shown as the angular structure 
of the kinetic-energy per solid-angle $dE_k/d\Omega$. The simulations that come closest to the structure identified here for $\Gamma(\theta)$ are those 
of A. Mizuta \& M. Aloy (2009), showing a power-law angular index between 2.1 and 2.5 (fig 7), but a similar value can be guessed for the Envelope's 
Lorentz factor structure (fig 5). 

 A low power-law index 1.5 is inferred by G. Urrutia et al (2023) for the structure of $dE_k/d\Omega$ for pressure-dominated jets, while a higher index 
2.5 is found for kinetic-dominated jets (fig 5). 
 These correlations between the mechanism accelerating jets and the jet properties suggest a $\Gc-g$ anticorrelation and a $\thc-g$ correlation, 
if the power-law angular dependences of the Lorentz factor $\Gamma(\theta)$ and kinetic-energy per solid-angle $dE_k/d\Omega(\theta)$ were the same.
 However, the $LAE$ temporal milestones depend only on the composite parameter $\Gctc$ and fits with that model to real light-curves cannot break the degeneracy, 
thus those correlations cannot be verified here.


\vspace*{1mm}
\subsection{\bf Indices of Envelope emission-characteristics distributions}

 Even less constrained by afterglow observations are the two emission characteristic indices $e$ and $i$ (panels D1 and D2 of Figures A3-A9) for 
the break-energy, $\Ep' \sim \theta^e$, and intensity at that break, $\Ip' \sim \theta^i$.
 Their large uncertainties (width $\Delta (e,i) > 2$ for the allowed ranges) indicate that afterglow optical and X-ray observations did not set 
the expected four stringent constraints on the asymptotic flux decays. 

 In many cases, that failure can be attributed to our requirement of a good optical monitoring, which introduces a selection effect in favor of 
the brightest optical afterglows, which in turn requires that the spectral-break between optical and X-ray is located close to the optical just 
after the Plateau begins, falling soon below the optical and completely coupling the optical and X-ray light-curves $\app^{(o)} = \app^{(x)}$ 
after the Plateau, thus reducing to three the number of observational constraints on four fundamental model parameters.

 An angle-indepedent (uniform) break-energy $\Ep'$ or intensity $\Ip'$ in the Envelope is possible in all cases, but not simultaneously. 
However, such an uniformity of only one break characteristic is not plausible. Instead, a variation of both could be suspected if the Envelope
has a different Lorentz factor than the Core. 

\vspace*{1mm}
\subsection{\bf Core and Envelope magnetic-field orientation} 

 Figure 2 shows that {\sl a parallel magnetic field induces a larger light-curve curvature $\Delta \alpha$ than an isotropic/perpendicular $B$} 
for both phases, with that curvature defined as the difference between the flux asymptotic power-law decay indices $\alpha$ (Eqs \ref{Bpar} and 
\ref{Bper}) at the end and start of each phase. 

 We note that the break-energy $\Ep$ is expected to fall from hard X-rays and cross the Swift/XRT 0.3-10 keV band during the prompt phase, 
which decouples the two XRT light-curves and leads to different curvatures $\Dagrb = \at - \ag$ at those two X-ray photon-energies. 
 That break is also expected to fall between optical and X-ray during the afterglow phase, leading to different curvatures $\Daag = \app - \ap$
in the optical and in the X-rays.

 Determining the $B$ orientation from the light-curve curvature comes with five caveats (\S\ref{issue1},\ref{issue2},\ref{issue345}) and a 
case-by-case analysis of those issues shows that our $B$ direction determinations are not weakened by the identified caveats.

 Panels E of Figures A3-A9 show that the {\bf Core magnetic field $B$} is found to be either {\bf perpendicular} or {\bf isotropic},
but a parallel magnetic field is never identified.
Conversely put, the prompt light-curve curvature $\Dagrb$ is smaller than expected for a parallel $B$. 
 A perpendicular $B$ is expected 
$i)$ for magnetically-dominated outflows advected from the ejecta source (H. Spruit, F. Daigne \& G. Drenkahn 2001), 
$ii)$ for the post-shock magnetic field generated by the two-stream instability (M. Medvedev \& A. Loeb 1999), 
$iii)$ from the differential compression of an ambient medium field by the forward-shock, 
and is most efficient at deflecting particles in the downstream region and at accelerating them, in the first-order Fermi acceleration mechanism. 
 An isotropic field (as a special type of random $B$) could be expected for magnetic fields produced by turbulence in the post-shock fluid.

 Panels E of Figures A3-A9 also show that {\bf parallel magnetic field $B$ in the Envelope} is identified for three of the seven afterglows 
fit here, meaning that, in some cases, the afterglow light-curve curvature $\Daag$ is larger than what a perpendicular $B$ can accommodate, 
given the best-fit values of the remaining three Envelope indices ($g;e,i$). Because of that extra-dependence, the Envelope $B$ orientation 
cannot be determined for four out of seven afterglows. 
 Note that an unconstrained $B$ orientation does not mean an isotropic field, whose signature (a time-independent flux-correction factor $X(\psi')$ -- 
Eq \ref{Xcorr}) is a dip in the $\chinu(\psi')$ curves shown in panels E at the magic angle $\Psi' = \arctan \sqrt{2} \simeq 55^\deg$. 

 For all three afterglows where the Envelope $B$-orientation (parallel) could be determined, we found a different orientation (perpendicular or isotropic)
than in the Core. That feature cannot be (cor)related to the Core being more relativistic than the Envelope because both are highly relativistic. 
We note (to no effect) that there are other differential features between those two outflow regions: the spectral characteristics break-energy $\Ep'$, 
break-intensity $\Ip'$, and high-energy spectral slope $\bH$. 

 The weaker frozen-in parallel magnetic field expected at large distances from the outflow origin and that such a field does not deflect particles 
efficiently (for first-order Fermi acceleration) suggest that these occasional identifications of a parallel $B$ from the afterglow light-curve curvature 
$\Daag$ may spurious and that, with the afterglow light-curve break $\Daag$ being set by three other parameters, the $B$-orientation in the Envelope 
may not be constrainable using afterglow observations.

\acknowledgments{ {\sl Acknowledgments}:
  This work made use of data supplied by the UK Swift Science Data Centre at the University of Leicester:
  {\sl www.swift.ac.uk/burst\_analyser} (P. Evans et al 2010). }



\begin{figure}
\vspace*{-1cm}
\centerline{\includegraphics[width=15cm,height=12cm]{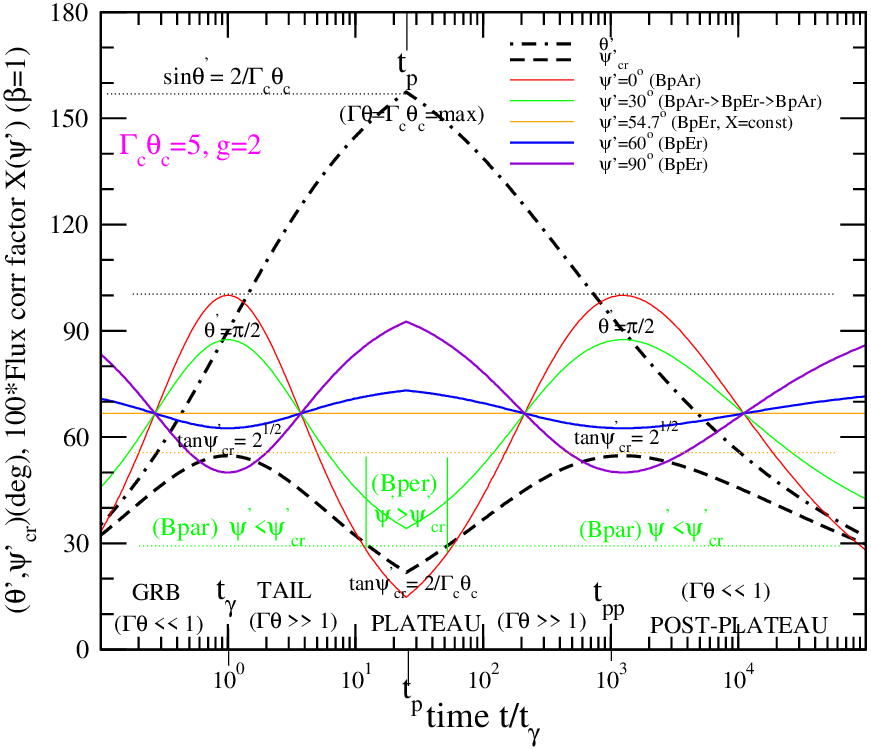}}
\figcaption{ 
  {\bf Comoving-frame emission angle} $\theta'$ (dot-dashed black) that is beamed in the direction of the observer increases from 0 at $t_o$ 
  (when the first photons arrive at observer) until epoch $\tp$ (end of Tail, beginning of Plateau) corresponding to the lowest relativistic radiation beaming 
  (maximal $\Gt=\Gctc$), after which it decreases to zero. 
  The GRB end/Tail start at $\tg$ and the Plateau end/post-Plateau beginning at $\tpp$ are both defined by $\Gt = 1$ and correspond to the observer exiting and 
  reentering, respectively, the cone of opening $\Gamma^{-1}$ of maximum relativistic beaming. From Equation (\ref{thprime}), it follows that, at those two
  milestones, $\theta' = \pi/2$. 
 {\bf Flux-correction factor} $X(\psi')$ for an unidirectional magnetic field oriented at fixed angle $\theta'$ relative to the local radial direction of flow, 
  for a Core parameter $\Gctc = 5$ and Envelope angular distribution of Lorentz factor $\Gamma(\theta) \sim \theta^{-g}$ with $g=2$ (yields a flat Plateau), 
  and for a synchrotron spectrum $F_\eps \sim \eps^{-\beta}$ with $\beta=1$ (yields simple analytical results and applies to X-ray afterglows).
  {\bf Critical angle $\psicr$}: 
  The factor $X(t,\psi') = X_{pAr} \cos^2\psi' + X_{pEr} \sin^2\psi'$ (Eq \ref{Xcorr}) is dominated by the parallel-$B$ factor for orientation angles 
  $\psi' < \psicr$ and by the perpendicular-$B$ factor for $\psi' > \psicr$, with the critical angle $\psicr$ (dashed black) given in Equation (\ref{psicr}). 
  For times of interest, the angle $\psicr$ ranges from a minimal value $\psi'_{cr,min} = \arctan [2/(\Gctc)]$ ($\Gctc \gg 1$) to a maximal 
  $\psi'_{cr,max} = \arctan \sqrt{2} = 55^\deg$. 
  For $\psi' \in (\psi'_{cr,min}, \psi'_{cr,max})$, the critical angle $\psicr$ crosses the fixed $B$-angle $\psi'$ at four epochs, when $X(\psi')$ 
  transits bewteen parallel and perpendicular $B$ "regimes".
   Using Equations (\ref{Xcorr}) and (\ref{psicr}), it can be shown that the {\bf extrema} of $X(t)$ and $\psicr(t)$ are reached at the temporal milestones 
   $\tg, \tpp$ (when $dX/dt=0$), and at $\tp$ (when $dX/dt$ changes sign).  
   The largest variation of $X(\psi')$ is obtained for a parallel $B$ ($\psi'=0$) (red curve); the amplitude of $X(\psi')$ variation decreases monotonically
  to zero at the magic angle $\Psi' = \arctan \sqrt{2}$, for which $X(t)=2/3$ is constant (orange), just as for an {\bf isotropic/omnidirectional $B$}.
  For $\psi' > \Psi'$, the amplitude of $X(\psi')$ variation increases monotonically with $\psi'$ but, even for $\psi'=90^\deg$ (perpendicular $B$), 
  it remains smaller than for a parallel $B$.
   By rearranging Equation (\ref{Xcorr}), it can be shown that the $X(t)$ correction factor has four {\bf nodes} when $\sin \theta' = \sqrt{2/3}$, 
  i.e. for $\theta'_{node} = \arctan \sqrt{2}$, where $X(t_{node})=2/3$ is independent of the $B$-angle $\psi'$. 
  The epochs of those four nodes, one for each GRB/afterglow phase, can be calculated using Equation (\ref{thprime}) for any $\Gt$ value.
  Those nodes will also be displayed at same epochs by the received flux.
}
\end{figure}

\begin{figure}
\vspace*{-1cm}
\centerline{\includegraphics[width=18cm,height=9cm]{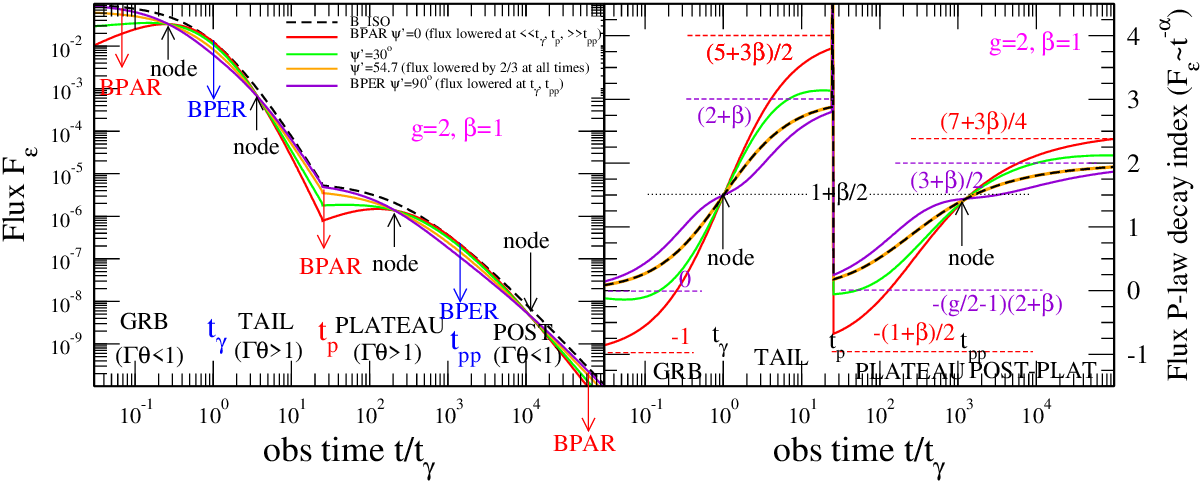}}
\figcaption{ 
 {\bf Left panel}: $LAE$ {\bf light-curves} for an Envelope Lorentz factor angular distribution $\Gamma(\theta) \sim \theta^{-g}$ with $g=2$,
  a synchrotron spectrum $F_\eps \sim \eps^{-\beta}$ with $\beta=1$, and comoving-frame emission characteristics break-energy $\Ep'$ and
  break-intensity $\Ip'$ with an uniform angular distribution in the Envelope (and Core).
  A parallel $B$ ($\psi'=0$ -- red line) has the strongest effect on the light-curve, by decreasing the flux at the emission onset $\to$, 
  at the Tail end/Plateau beginning $\tp$, and well after the post-Plateau ($t \gg \tpp$).
  Non-parallel magnetic fields ($\psi'=90^\deg$ -- blue line) have a weaker effect, by decreasing the flux at the GRB end/Tail beginning $\tg$ 
  and at the Plateau end/post-Plateau beginning $\tpp$.
  Equation (\ref{FXpsi}) shows that {\bf nodes of the received flux} should occur at the nodes for the flux-correction factor $X(t,\psi')$ of Fig 1, 
  where that factor is independent of the magnetic field orientation angle $\psi'$.
 {\bf Right panel}: Flux power-law decay index (logarithmic derivatives) $\alpha = - d\log F_\eps/d\log t$ calculated numerically and compared 
  with the analytical expectations (Eqs \ref{Bpar} and \ref{Bper}) for the asymptotic values (horizonthal dashed lines).
  Equation (\ref{FXpsi}) indicates that the {\bf nodes of the decay index} (when $\alpha$ is independent of the angle $\psi'$) should occur 
  when the flux-correction factor satisfies $dX/dt=0$, i.e. at two of $X(t)$ extrema epochs: end of GRB ($\tg$) and end of Plateau phase ($\tpp$). 
  For this particular choice of index $g=2$ and slope $\beta=1$, the decay index $\alpha = 1 + \beta/2$ at both nodes is the average of the asymptotic 
  indices at the Tail end and Plateau beginning. 
  The numerical decay indices approach the expected asymptotic values by $\delta \alpha \simeq 0.25$ at 1.5 dex in time after/before the epoch 
  of those corresponding node, which shows why numerical fits are more powerful than using analytical asymptotic indices, especially for afterglow 
  phases that are monitored less than 2 decades in time.
  A parallel $B$ leads to a faster GRB rise, a faster decaying tail, a faster rising Plateau, and a faster decreasing post-Plateau than a perpendicular $B$,
 and {\bf increases the curvature of the prompt/GRB and delayed/afterglow light-curves}, where "curvature" is defined as the difference between 
 the asymptotic flux-decay indices.
}
\end{figure}

\clearpage

\appendix 

\section{\bf Numerical Fits}

 Figures A3-A9 shows the best-fits obtained with the $LAE$ model for {\bf seven} Swift/XRT GRBs and afterglows. 
These bursts were chosen because they have all four X-ray phases well-monitored and a good coverage of the optical afterglow,
both of which are pre-requisites for a good chance of determining the five fundamental $LAE$ model parameters.
 All but two light-curves fit in this work are from the first 1.5 years of Swift observations. If optical coverage of subsequent GRBs were as good 
as for those five "initial" bursts, then $LAE$ fits could be applied to another five dozen GRBs.

 A general description of those fits and figures is given below. Specific details are presented in the figure captions.

\vspace*{1mm}
\subsection{\bf A1. Light-Curve Fits (Upper Panel)} 
\label{issue1}

 Upper panels show the $\sim 3$ eV optical (prompt/brown and delayed/black) and X-ray flux (10 keV/dark-green and 0.3 keV/orange) for the burst 
and afterglow emissions, with the {\sl prompt} X-ray {\sl specific} flux calculated from the XRT 0.3-10 keV {\sl bolometric} flux using a 
{\sl broken power-law} spectrum with the low-/high-energy asymptotic slopes ($\bL,\bH$) and break-energies $\Ep$ listed or shown in panel A,  
and with the {\sl delayed} flux calculated for a {\sl single power-law} spectrum with the high-energy slope $\bH$ shown in panel B.
 Light-green points show the 10 keV flux calculated from the 15-50 keV BAT flux for the spectral parameters of panel A, but the BAT data 
are not included in the fit even when they match well the XRT measurements (see below).

 Observer's time $t$ measured since GRB trigger is shifted closer to the instantaneous emission-epoch $\to$ to show better the prompt phase
and its flux power-law decay.
 For illustration, XRT X-ray data within (3-5)\% of their measurement epoch (since trigger) are rebinned and the measurement uncertainty is 
calculated by weighting the errors of individual measurements. For model fitting, the XRT data are used without rebinning.


 Best-fits obtained with the $LAE$ model by $\chi^2$ minimization are shown for two orientations of the magnetic field $B$: 
{\sl parallel} to the flow direction (red lines or symbols) and {\sl perpendicular} to it (blue lines or symbols).
 The best-fit reduced $\chinu = \chi^2$/dof (degrees of freedom = number of data minus number of free model parameters) is always sufficiently 
large to rule out the best-fit because the data have statistically-significant fluctuations that cannot be captured by the $LAE$ model, 
where smooth angular distributions of the dynamical and emission parameters lead to smooth light-curves. 

 Discontinuities in the 0.3 keV and 10 keV model fluxes at the Tail-Plateau transition are due to discontinuities at the Core-Envelope boundary
allowed for all three emission characteristics high-energy slope $\bL$, break-energy $\Ep'$ and break-intensity $\Ip'$.  
That first first discontinuity is required by the effective spectral slope $\bx$ reported by Swift/XRT (panel B) and is not always sudden, 
with many GRBs displaying a fast hardening at the end of the Tail.
 That spectral hardening is not captured by the $LAE$ model which assumes that spectral slopes are constant. 

 {\bf Issue \#1:} {\sl Effect of $\to$ on the determination of the magnetic field orientation $\psi'$ in the Core}.
 An important factor for determining the angle $\psi'$ is the epoch $\to$ when emission is released. 
That is due to that a parallel magnetic field ($\psi'= 0$) yields a fast rising light-curve followed by a decay, while a perpendicular field 
($\psi'= 90^\deg$) yields only decaying light-curves (first lines of Eqs \ref{Bpar} and \ref{Bper}). This means that a $\to$ prior to that onset 
could favor a parallel $B$ (owing to its fast flux-rise), while a $\to$ later than the decay-onset could favor a perpendicular field (owing to
its pre-existing flux-decay). 

 Consequently, it is safer to set a fixed $\to$ at the onset of the prompt flux decay but, given the "stiffness" of the prompt light-curves 
(first two lines of Eqs \ref{Bpar} and \ref{Bper}), there will always be some suspicion that the choice of $\to$ introduces an unwanted bias toward 
a certain $B$ orientation. However, letting the emission-epoch $\to$ be a free parameter could lead to a late $\to$ that would exclude many
early measurements that cannot be accommodated by a perpendicular $B$ (which yields only decaying light-curves), and that would give an unfair 
advantage to that orientation.

 The best solution is to adjust a fixed $\to$ (within a reasonable range) and assess the robustness of the $B$ field orientation determination.
Another rationale for adjusting $\to$ appears when the XRT has missed the pulse peak, in which case the BAT data offers guidance as to where
$\to$ could be.

\vspace*{1mm}
\subsection{\bf A2. Break-Energy $\Ep$ (Panel A)} 
\label{issue2}

 Green and black symbols are break-energies $\Ep$ in (15,50) keV and (0.3,10) keV, respectively, for the Core prompt emission, calculated for 
the asymptotic low and high-energy slopes $\bL$ and $\bH$ indicated in panel B, by requiring that the resulting hardness-ratio (25-50k/15-25k
photon count ratio for BAT and 1.5-10k/0.3-1.5k ratio for XRT) is the same as for a power-law spectrum with the effective slope $\bx$ reported 
by BAT or XRT for the unabsorbed emission, obtained after correction for known Galactic dust-absorption and for the host dust-absorption determined 
from the afterglow hardness-ratios (P. Evans et al 2010), i.e. at times sufficiently late that the break-energy $\Ep$ has fallen below the XRT
band and when it is safe to assume that the XRT continuum is a single power-law (N. Butler \& D. Kocevski 2007). 

 The {\sl XRT break-energies $\Ep$ calculated here are in agreement with detailed fits} to the XRT+BAT spectrum (N. Butler \& D. Kocevski 2007) 
for GRBs 060614, 060729, 061121 (Figures A3-A5), but a factor $\siml 2$ below those of GRB 060526 (Figure A7). With the inferred break-energy $\Ep$ 
and the assumed spectral slopes $\bL$ and $\bH$, we calculate the flux $\Fp$ at the break-energy by requiring that the corresponding 15-50 keV 
and 0.3-10 keV band fluxes match the unabsorbed fluxes reported by Swift/BAT and XRT, respectively. 
The equations used to calculate $\Ep$ and $\Ip$ and their uncertainties are given in Appendix A of an article by A. Panaitescu (2025).
  

 Black lines in panels A show the break-energies $\Ep$ of the best-fit $LAE$ model, only for comparison with the data: the indirectly measured 
break-energies $\Ep$ are not fit, with the model fits using only specific fluxes.

 For any reasonable choice of asymptotic spectral slopes, there is always a discrepancy (of a factor up to 10) between the break-energies $\Ep$ 
calculated from BAT and XRT effective spectral slopes $\bx$, even when BAT and XRT measurements are simultaneous.
 Given the above-mentioned agreements between our XRT $\Ep$ "measurements" and values determined from actual spectral fits, we conclude that 
our calculations of $\Ep$ from BAT measurements is erroneous, most likely because the spectral softening displayed by the BAT continuum (panel B) 
is due to a softening of the high-energy slope $\bH$ above the break-energy $\Ep$ that has already fallen below the BAT range (as indicated by 
the $\Ep$ inferred from XRT data).

 For that reason, BAT 10 keV light-curves are not included in model fitting even in those rare instances when XRT has missed the pulse peak,
even if the calculated BAT 10 keV fluxes are compatible with those inferred from the XRT data after the missed XRT light-curve peak.

 However, if that BAT-XRT $\Ep$ discrepancy were real, then we would conclude that there are two spectral breaks in the XRT and BAT domains 
during the prompt phase. 
 G. Oganesyan et al (2017) have shown that there are two breaks in the continuum of a dozen Swift/XRT+BAT and Fermi/GBM spectra,
with the a lower-energy break in the XRT window and a higher-energy break starting above the BAT range and decreasing into it in five cases. 
From the spectral slopes below and above the lower-energy break, they identify that break with the cooling-break $\Ec$, above which radiate 
electrons that cool on a timescale shorter than their age.
 Then, BAT+XRT measurements can be used to choose wisely the spectral slopes above and below each break (the cooling-break $\Ec$ or 
the injection-break $\Ei$ at which the electrons radiate initially). 

{\bf Issue \#2:} {\sl Effect of $\Ep$ on the determination of $\psi'$ in the Core}.
As discussed in \S\ref{limits}, a measured break-energy $\Ep$ decreasing faster than the model expectation favors a parallel magnetic field. 
Conversely, an $\Ep$ decreasing slower than expected (which does not seem to ever happen) favors a perpendicular $B$.

\vspace*{1mm}
\subsection{\bf A3. X-ray Effective Spectral Slope $\bx$ (Panel B)} 
\label{issue345}

 The peak energy $\Ep$ decreases in time, sweeping from above a given observational window.
At earlier times the break-energy $\Ep$ is within or above the BAT range (blue) and above the XRT window (brown), thus the effective slope $\bx$ 
shows the low-energy slope $\bL$ of the broken power-law spectrum (B.-B. Zhang et al 2006). 
When $\Ep$ falls in the observing band, the effective slope $\bx \in (\bL,\bH)$. 
At later times the break-energy $\Ep$ is below the observational domain, and $\bx$ provides a harder limit on the high-energy slope $\bH$.
Thus, the {\sl asymptotic} effective slopes measured by BAT and XRT at early and late times provide constraints on two spectral slopes needed to calculate 
the break-energy $\Ep$ of the Core emission continuum by matching the measured hardness-ratio (or its equivalent, the effective spectral slope).

 In $LAE$ fits, the {\bf low-energy slope} $\bL$ is a fixed parameter and is allowed to have two values: $\bL=-1/3$ ($\bL=1/2$) for electrons 
that cool on a timescale longer (shorter) than the timescale on which they are accelerated. Early-time asymptotic effective spectral slopes 
$\bx$ measured by Swift are close to these two fixed values, but measurements of a softer $\bx \simeq 1/2$ do not necessarily require that 
$\bL=1/2$ because an effective spectral slope softer than $\bL=-1/3$ could result if the break-energy $\Ep$ is already in the observational 
domain at early times.

 Because the break-energy $\Ep$ is always below the X-rays during the afterglow phase, the low-energy slope $\bL$ for the Envelope emission 
affects only the optical flux. For numerical fits, both possible values for $\bL$ are allowed.
 The reason for this allowance is that the Core and Envelope electrons radiating at the peak of the synchrotron spectrum could be in different 
cooling regimes at different times, i.e. the spectral break $\Ep$ located between optical and X-ray during the afterglow phase could be either 
the "cooling-break" $\Ec$ (then $\bL=1/2$) or the "injection-break" $\Ei$ (then $\bL=-1/3$). 
 For half of the afterglows considered here, the Envelope emission spectral slope $\bL$ remains ambiguous, with a harder $\bL=-1/3$ and 
a correspondingly lower $\Ep$ providing the same fit quality to the optical light-curve as a softer $\bL=1/2$ and a higher $\Ep$.
We note that the fixed choice of $\bL$ has an effect on the fundamental parameters ($g;\psi';e,i$) because the afterglow optical flux decay
(indices $\ap$ and $\app$ in Eqs \ref{Bpar} and \ref{Bper}) depends also on $\bL$.

 The {\bf high-energy slope} $\bH$ is set by the non-universal (R. Shen, P. Kumar, E. Robinson 2006) electron power-law distribution with energy.
Because much rides on the slope $\bH$ for the Core emission (the Tail decay, the 0.3k-to-10k flux-ratio), this parameter cannot be fixed 
at the late-time asymptotic effective slope $\bx$ measured by XRT during the Tail, which may set only a lower limit on $\bH$ if measurements
do not extend until the break-energy $\Ep$ has fallen below the observing window.

 The Envelope emission slope $\bH$ sets the afterglow decay, the optical-to-X-ray flux ratio, and the 0.3-to-10k flux-ratio. 
For that reason, the Envelope spectral slope $\bH$ is also a free parameter.

 Black lines in panels B show the XRT effective slope $\bx$ for the $LAE$ best-fit, calculated in the same way as the effective slopes published 
by the XRT team, as the single power-law spectrum that yields the same (1.5-10)k/(0.3-1.5)k photon-flux hardness-ratio as the broken power-law 
spectrum with the fixed slope $\bL$, best-fit slope $\bH$, and break-energy $\Ep$ (shown in panel A).
This comparison between measured and model effective spectral slopes is just for illustration: the measured slope $\bx$ is not fit and the model-fits 
use only fluxes.


{\bf Issue \#3:} {\sl Effect of $\bL$ on the determination of $\psi'$ in the Core}.
 In those cases where the break-energy $\Ep$ is not above, but in the observing window at the earliest measurements, the measured early-time asymptotic 
slope $\bx$ during the prompt phase sets only a soft limit on the actual low-energy slope $\bL$ of the Core emission. Then, a soft measured slope 
$\bx = 1/2$ allows for a harder $\bL = -1/3$. 
 Equations (\ref{Bpar}) and (\ref{Bper}) show that the expected curvature of the prompt emission light-curve below $\Ep$ (i.e. at 0.3 keV) 
is $\Dagrb^{(pAr)} = 2\bL + 3$ and $\Dagrb^{(pEr)} = \bL + 2$. This means that a harder low-energy slope $\bL = -1/3$ yields a smaller
curvature of the $LAE$ 0.3 keV light-curve than a softer $\bL = 1/2$ and, to compensate for that loss of curvature, the $LAE$ model will "move" 
toward a parallel $B$.
Thus, the possibility of {\sl a harder low-energy slope $\bL$ half-biases the $LAE$ fit toward a parallel $B$} to accommodate the curvature 
of the XRT 0.3 keV prompt light-curve, 

{\bf Issue \#4:} {\sl Effect of $\bH$ on the determination of $\psi'$ in the Core}.
 In a similar way, the opposite bias happens for the prompt phase emission when considering a softer high-energy slope $\bH$.
In those cases where the break-energy $\Ep$ is not below, but in the observing window at the latest measurements (which may happen when the Earth's
occultation occurred before the Tail end), the measured late-time asymptotic slope $\bx$ during the prompt phase sets only a hard limit on the 
high-energy slope $\bH$ of the Core emission and motivates one to assume a softer slope $\bH$ which would increase the prompt light-curve curvature
above $\Ep$ (i.e. at 10 keV).
Thus, the possibility of {\sl a softer high-energy slope $\bH$ half-biases the $LAE$ fit toward a perpendicular $B$}, to account for the measured 
curvature of the 10 keV light-curve.

{\bf Issue \#5:} {\sl Effect of the Tail spectral hardening on the determination of $\psi'$ in the Core}.
 The gradual spectral hardening displayed by some GRB Tails is most likely attributable to a hardening of the Core's high-energy slope $\bH$,
and that mitigates the 10 keV light-curve decay because the break-energy $\Ep$ is in the XRT window. As discussed in \S\ref{limits}, a smaller
curvature of only one of the two prompt light-curves {\sl introduces a half-bias of the $LAE$ model toward a perpendicular $B$}.


\vspace*{1mm}
\subsection{\bf A4. Basic Model Parameters (Panels C-E)} 

 The {\bf prompt emission} does all of the following: \\
(1) {\sl determines} the Core dynamical parameter $\Gctc$ (panel C1 - black line) through the timing of the GRB $(\Gctc)^2 = \tp/\tg$ and \\
(2) the Core B-field angle $\psi'$ (panel E - black line) through the curvature of the GRB light-curve $\Dagrb = \at -\ag$, and \\
(3) {\sl constrains} the Envelope dynamical parameter $g$ that sets the Plateau dynamical range 
    $\tpp/\tp = 2(\Gctc)^{2/(g-1)}$ (Eq \ref{Gt}) through the Core parameter $\Gctc$.

 The {\bf delayed/afterglow emission} (red and green lines) {\sl determines} \\
(1) the Core dynamical parameter $\Gctc$ through the afterglow milestones: $\Gctc = (\tpp/2\tp)^{(g-1)/2}$. \\
(2) the Envelope dynamical parameter $g$ (panel C2) through the afterglow milestones: $g = 1 + 2\ln (2\Gctc)/\ln (\tpp/\tp)$, \\
(3) the Envelope emission parameters $e,i$ (panels D1 and D2) through two linear combinations $i+e\beta$ 
     (with $\beta=\bo$ or $\beta=\bx$) and \\
(4) the Envelope B-field orientation angle $\psi'$ (panel E), \\ 
through the four asymptotic flux-decay indices $(\ap,\app)^{(o,x)}$.

 For all GRB light-curves fit here, the best-fit $\chinu$ is larger than unity. In a few cases, that can be identified to a systematic discrepancy
between data and model but, in most cases, that large $\chinu$  is due to X-ray fluctuations that are statistically significant but which 
cannot be accommodated by the $LAE$ model, which yields only smooth light-curves.
 
 For the joint variation of all 13 $LAE$ model parameters, the $2\sigma$ confidence level of parameter corresponds to a $\chi^2$ variation
of $\simeq 24$ around the minimal $\chi^2$. In this work, most light-curves have 400-1000 measurements, thus the $2\sigma$ confidence level 
corresponds to variation of the reduced $\chi^2$ of $\Delta \chinu \simeq 0.025-0.06$. Because $\chinu$ curves for a single parameter of interest 
(panels C-E) are not sufficiently well-sampled to allow reading of such a small $\chinu$ variation, we estimate allowed parameter ranges 
using a more generous $\Delta \chinu \simeq 0.1$ variation around the minimal $\chinu$. A second reason for increasing $\Delta \chinu$ is 
a suspicion that, if the best-fit is statistically impossible (because of a $\chinu$ above unity, for a very large number of degrees of freedom), 
then using the standard variation of $\chinu$ arounds its minimum could lead to artificially small allowed parameter ranges, with the ad-hoc
rule-of-thumb that, if the best-fit has $\chinu = 2-3$ (times larger than what is statistically acceptable), then more realistic confidence 
intervals should use $(2-3)\Delta \chinu$ for determining the allowed parameter ranges.

 We expect that a good optical coverage of the Plateau and post-Plateau will yield sufficient constraints to determine all basic Envelope parameters 
($g;e,i;\psi'$).
 To understand the relative importance of optical and X-ray data, $\chinu(\psi')$ curves of panels C-E are shown for all afterglow observations
(red lines) and for only the more numerous X-ray data (green lines).
 Then we continue by adding the less numerous prompt measurements (black lines), expecting a tight determination of the Core parameter $\Gctc$.

 Figures A3-A9 show that, in general, afterglow observational alone do not constrain well the Core parameter $\Gctc$, and that prompt measurements
restrict the allowed range to $\Delta \Gctc \siml 1$ in about half the GRBs fit here. Furthermore, afterglow measurements constrain the
Lorentz factor index also for half of GRBs by reducing the width of the allowed range to $\Delta g < 1$, most allowed values being around
$g \siml 2$, the value expected for flat Plateaus. Finally, afterglow measurements only rarely determine (constrain well)the Envelope emission 
characteristics indices $e$ and $i$, with allowed ranges often extending over $\Delta (e,i) > 2$, contrary to our initial expectations.


\vspace*{1mm}
\subsection{\bf A5. Magnetic-Field Orientation (Panel E)} 

 The B-field orientation angle $\psi'$ changes the curvature (quantified by the difference between asymptotic power-law flux-decay indices $\Da$) 
of the prompt and afterglow light-curves, with parallel fields ($\psi'=0$) inducing more curvature than perpendicular ones ($\psi'=90^\deg$). 
 For the prompt emission, $\Dagrb (\psi') = \at (\psi') -\ag (\psi')$ depends on the spectral slope $\bx$; for the afterglow emission, 
$\Daag (\psi') = \app(\psi') -\ap(\psi')$ is also set by the Lorentz factor angular index $g$ and the linear combination $i+\bx e$ 
of emission-characteristics angular indices.

 We start the determination of the $B$-angle $\psi'$ in the Envelope with the more numerous afterglow measurements (red lines) and add the fewer
prompt measurements (black lines) to identify the orientation angle $\psi'$ in the Core. 
 To interpret correctly the $\chinu$ curves shown in panels E, one should keep in mind the following: \\
(1) the curvature of afterglow light-curves is determined not only by the $B$-field orientation angle $\psi'$ but also by three other Envelope 
   fundamental parameters: Lorentz factor $\Gamma \sim \theta^{-g}$, break-energy $\Ep' \sim \theta^e$, and break-energy $\Ip' \sim \theta^i$ 
   angular distribution indices ($g,e,i$), as shown by Equations (\ref{Bpar}) and (\ref{Bper}). Thus, a flat $\chinu (\psi')$ curve does not imply
   an isotropic magnetic field but that, for any fixed orientation, the remaining three parameters ($g,e,i$) are adjusted to yield a light-curve 
   whose curvature matches the observations. In fact, an isotropic field, for which the flux-correction factor $X(t,\psi')$ is time-independent, 
   would be identified by a dip in the $\chinu (\psi')$ curve at the magic angle $\Psi' = \arctan \sqrt{2}$. \\
(2) the flux-correction factor $X(\psi')$ of Equation (\ref{Xcorr}) has a stronger variation with time for a parallel $B$ (or small angle $\psi'$) 
   than for a perpendicular $B$ (or largish angle $\psi'$), which is illustrated in Figure 2 by the weaker variation of $X(\psi')$ for $\psi > 45^\deg$, 
   thus, to first order, matching the light-curve curvature means discriminating between a parallel and a perpendicular/isotropic magnetic field, 
   in the sense that $\chinu (\psi')$ should be flat(tish) at larger angles $\psi'$. \\
(3) owing to the uniform Core assumption, the curvature of the prompt emission light-curves is determined by the $B$ orientation and by high and 
   low-energy spectral slope $\bL$ and $\bH$. That may suggest that a robust determination of the magnetic field orientation is more feasible for
   the Core than for the Envelope, but there are {\bf five issues} that weaken the former.. 
   Three issues are due to possible uncertainties in the emission-epoch $\to$ and the asymptotic effective slope $\bx$, which sets the low and 
   high-energy slopes $\bL$and  $\bH$.
   Two issues arise from the limitations of the $LAE$ model (in its current version) to accommodate ubiquitous features in the prompt emission: 
   a faster-than-expected decrease of the break-energy $\Ep$, and a spectral hardening at the end of Tail.
   Elucidating these issues requires further fits, as discussed in the figure captions, for each GRB. \\
(4) $\chinu (\psi')$ is calculated for afterglow data alone and for the full data set, including prompt measurements, thus a change in the $B$ 
   orientation from the former data set to the latter means that the Envelope and Core magnetic fields are perpendicular to each other.  
   
 With the above rules, Figures A3-A9 show that, whenever (3 out of 7 cases) afterglow light-curves determine the magnetic field orientation in 
the Envelope, that $B$ is parallel to the flow direction, and that the $B$ orientation in the Core required by prompt observations is equally 
often found to be perpendicular to the flow direction or isotropic.

\clearpage

\begin{figure}
\vspace*{-1cm}
\centerline{\includegraphics[width=15cm,height=16cm]{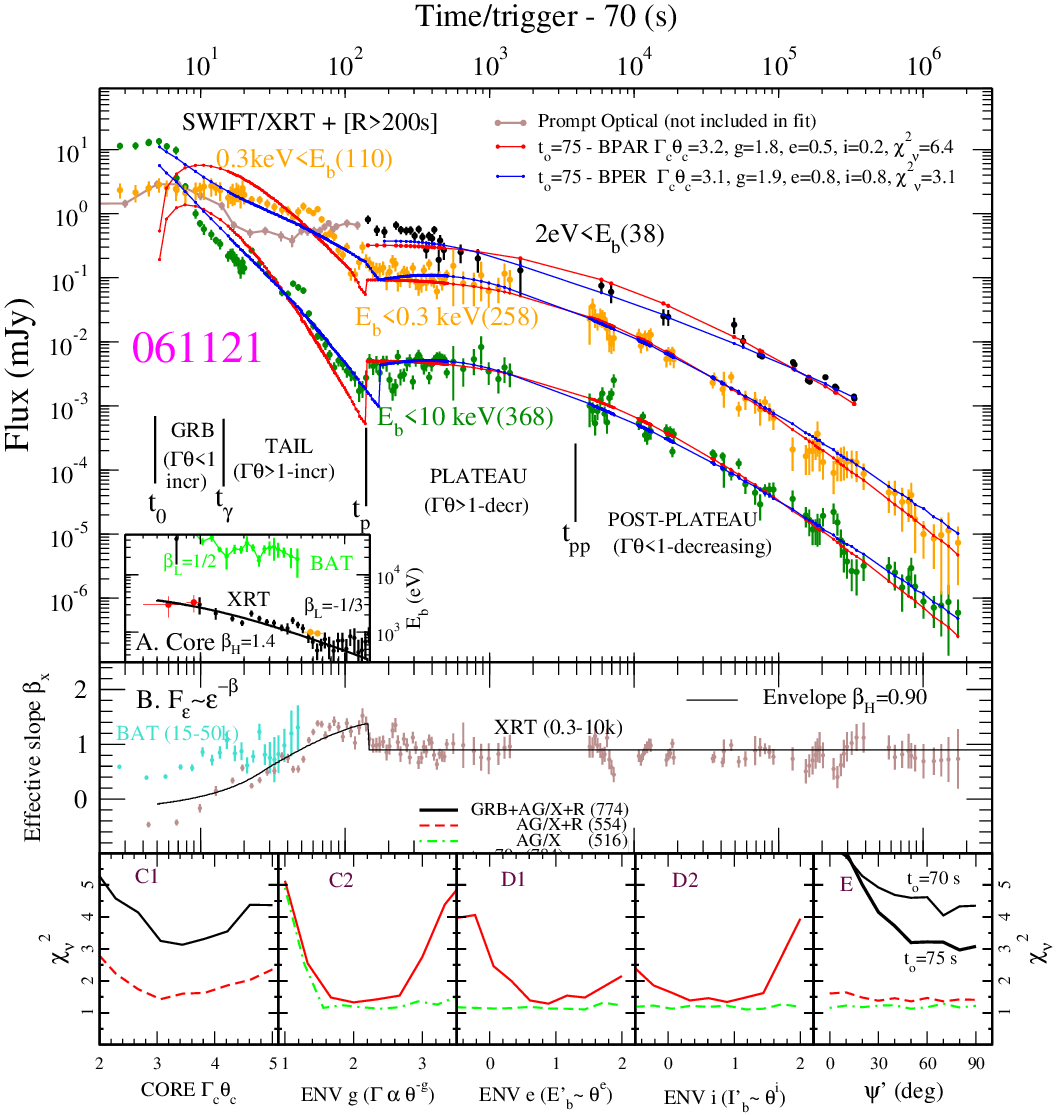}}
\figcaption{ {\bf GRB 061121}.
Upper panel: 
	Prompt ($t < \tp$) and delayed ($\tp < t$) optical (black symbols) and X-ray (10k/green, 0.3k/red) data vs $LAE$ model best-fits (parameters in legend), 
	for two orientations of the magnetic field (red symbols and lines for parallel $B$, blue for perpendicular) (number of each type of data is given in brackets).
	The instantaneous $LAE$ emission-epoch is $\to=75$s, when the 10 keV flux displays a peak. 
	{\bf The XRT has monitored the entire prompt phase}.
	The prompt optical data before 200 s are not included in the fit because they are not sufficiently well-coupled with the X-ray light-curve. 
	(The continuity of optical light-curve from prompt to afterglow indicate a single origin and require further modification of the $LAE$ model, 
	including a different break-energy above optical during the prompt phase than during the afterglow).
        Optical data from K. Page et al (2007) and T. Uehara et al (2011).
Panel {\bf A}: 
	BAT (green) and XRT (black) break-energies $\Ep$ calculated for a broken power-law spectrum from the reported effective spectral slopes $\bx$ (panel B). 
	Red (orange) symbols show earlier (later) the energy-breaks $\Ep$ at 1-few keV measured by G. Oganesyan et al (2017) (N. Butler \& D. Kocevski 2007) 
	from direct fits to the XRT spectrum. 
	Note the agreement between the $\Ep$ calculated expeditiously here (black symbols) from the XRT effective spectral slope and those obtained
	from detailed fits to XRT spectra.
	The best-fit $LAE$ model break-energy $\Ep$ (black line) is in agreement with measurements (black symbols).
Panel {\bf B}: 
	The gradual softening of the BAT effective spectral slope $\bx$ (cyan) leads to a break-energy (panel A) that decreases through the 15-50 keV BAT window. 
	However, this higher $\Ep$ may be just an artifact of the underlying assumption of a broken power-law spectrum of constant high-energy slope $\bH$.
	The gradual softening of the XRT effective spectral slope $\bx$ (brown) indicates a break-energy (panel A) that decreases through the 0.3-10 keV XRT window. 
	The {\bf early asymptotic XRT spectral slope} (brown)$\bx$ {\bf requires a low-energy slope} $\bL=-1/3$ (uncooled electrons) for the Core emission. 
	The optical is below $\Ep$ (upper panel) and requires a low-energy slope $\bL=-1/3$ for the Envelope emission because the slope $\bH=1/2$ 
        for cooled electrons would over-predict the optical $LAE$.
	The black line shows the XRT effective slope $\bx$ for the $LAE$ best-fit. 
Panel {\bf C1}: 
	The Core parameter $\Gctc$ is constrained by the temporal milestones ($\tp/\tg, \tpp/\tp$) of the prompt and delayed light-curves (upper panel). 
	Addition of the former (black line) to the latter (red) reduces the allowed range to $\Gctc \in (2.9,3.5)$.
	The large increase in $\chinu$ when the prompt data are added to the fit is due to the significant departure of the 10 keV light-curve from the broken 
        power-law of the $LAE$.
	Legend indicates in round brackets the number of measurements that were fit.
Panels {\bf C2-E}: 
	The break-energy $\Ep$ is between optical and X-ray during the entire afterglow phase (upper panel), thus the {\bf afterglow data should determine
	all four fundamental Envelope parameters}.
Panel {\bf C2}: 
	The Envelope's power-law angular structure of the Lorentz factor $\Gamma \sim \theta^{-g}$ has index $g \in (1.7,2.4)$.
Panels {\bf D1, D2}: 
	Envelope emission characteristics indices $E'_b \sim \theta^e$ - $e \in (0.6,1.4)$ and $I'_p \sim \theta^i$ - $i \in (0.3,1.2)$. 
Panel {\bf E}: 
	The afterglow light-curve curvatures are consistent with {\bf any magnetic field orientation in the Envelope} (flat red line), which means that other 
        $LAE$ model parameters ($g,e,i$) can be adjusted to increase or decrease the light-curve curvature induced by the $B$-field orientation.
	However, a parallel magnetic field yields a too fast decay and an excessive curvature of the prompt 0.3 and 10 keV fluxes,
	and best-fits to the prompt emission indicate that $\psi' > 50^\deg$, i.e. a {\bf perpendicular $B$ in the Core}.
	As expected, an earlier emission-epoch $\to = 70$s (thin line) disfavors a perpendicular $B$ because such an orientation yields only decaying light-curves, 
        which are incompatible with a pre-peak flat or rising flux, but that pertains only to a limitation of our treatment of the $LAE$ model.  
}
\end{figure}

\begin{figure}
\vspace*{-1cm}
\centerline{\includegraphics[width=15cm,height=16cm]{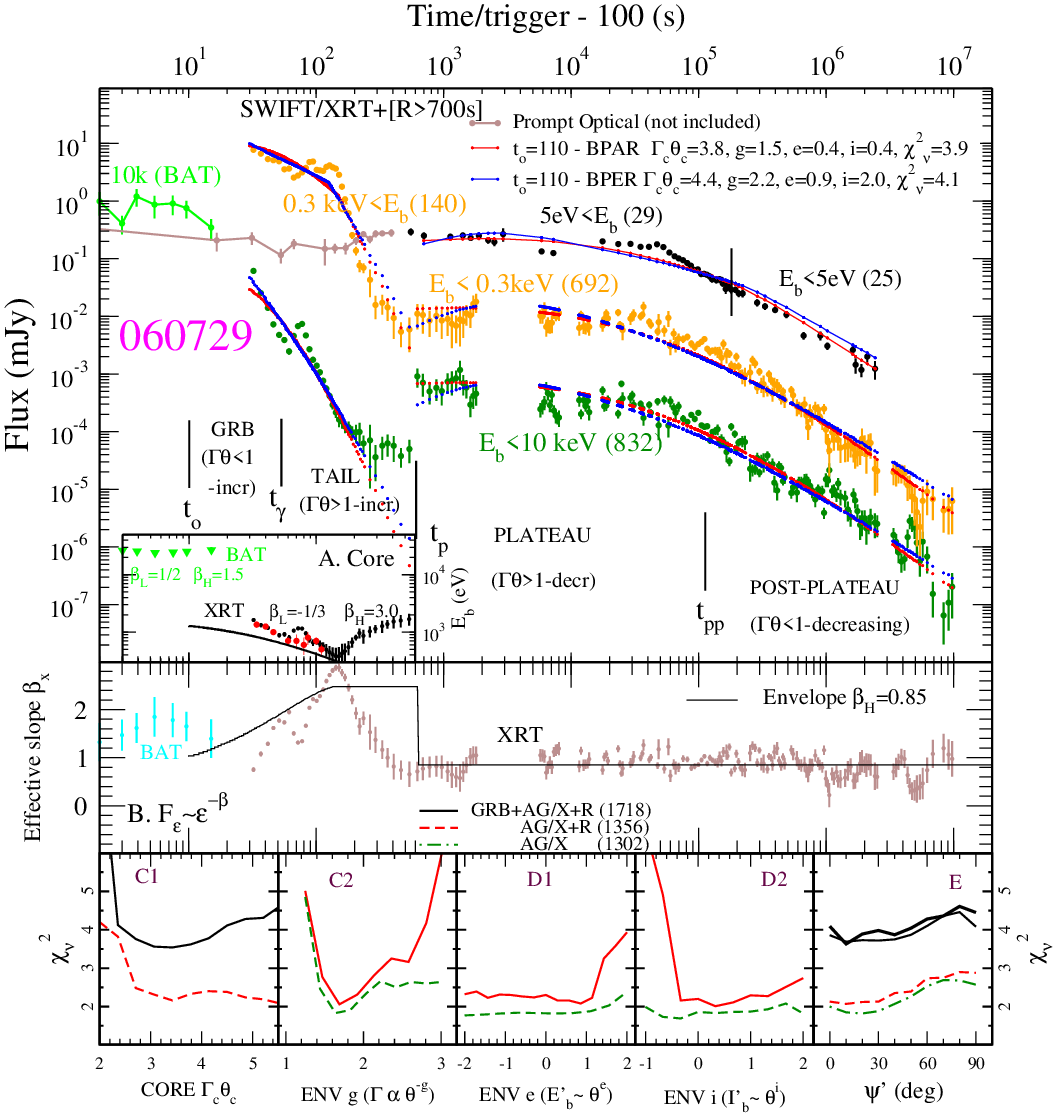}}
\figcaption{ {\bf GRB 060729}. 
Upper panel:
	The emission-epoch is set at $\to=110$ s, around the peak of the BAT 10 keV light-curve, but could be several seconds earlier.
	{\bf The XRT has monitored most of the prompt phase} (end of GRB and whole Tail).
	The 0.3 keV light-curve break at 240 s is due to the passage of the break-energy $\Ep$ (panel A).
	Only afterglow optical data after 700 s are included in the fit because the prompt optical emission is decoupled from the X-ray.
	(As for GRB 061121, the continuity of optical light-curve from prompt to afterglow implies a common origin and requires another spectral-break
	above the optical during the prompt phase).
	The optical data can be fit equally well with an Envelope spectral slope $\bL=-1/3$ (uncooled electrons) and $\bL=1/2$ (cooled electrons).
        Optical data from D. Grupe et al (2007)
Panel {\bf A}: 
        The BAT effective spectral slope (panel B) does not display a softening and the inferred break-energies should be considered only upper limits.
	Red symbols show the $\Ep$ measurements of N. Butler \& D. Kocevski (2007) from direct spectral fits. 
	Our speedy $\Ep$ determinations agree with those measurements. 
	The assumption of a fixed high-energy spectral slope $\bH$ at the end of the Tail leads to an increasing $\Ep$ during the spectral hardening 
	at 250-700 s (panel B), which leads to a lesser underestimation of the XRT 0.3 keV flux (because $\bL=-1/3$), a larger overestimation of 
	the XRT 10 keV flux (because $\bH=3.0$), and a decoupling of the X-ray light-curves at the Tail end.
	The best-fit model $\Ep$ under-predicts measurements and their decrease in time.
Panel {\bf B}: 
	The {\bf early-time asymptotic XRT slope allows both $\bL=1/2$ and $\bL=-1/3$}.
	The decreasing $\Ep$ leads to a softening of XRT slope until 250 s, when $\Ep$ should have fallen to 0.3 keV, and indicates a Core emission with a 
	high-energy slope $\bH=3.0$, followed by a strong softening of the XRT continuum before the end of the prompt phase. 
Panel {\bf C1}: 
	As expected, the Core parameter $\Gctc \in (2.6,4.3)$ is much better determined by the prompt emission (black line includes it) than by the afterglow (red). 
	The larger $\chinu$ of the former is due to fluctuations of the prompt emission.
Panels {\bf C2-E}: The break-energy $\Ep$ is between optical and X-ray only during the afterglow Plateau, thus one observational constraint (post-Plateau asymptotic
        flux-decay index at energy below $\Ep$) is missing, and the {\bf afterglow data may not constrain well the Envelope parameters}.
Panel {\bf C2}: 
	The index for the Envelope angular structure of the Lorentz factor $g \in (1.6,1.9)$ is well determined by the X-ray afterglow alone (green)
	and optical data do not narrow that range.
Panels {\bf D1, D2}:
	Broad allowed ranges for the indices for the Envelope spectral characteristics $e < 1.2$ and $i \in (-0.3,1.4)$ arise from above limitation of afterglow data.
Panel {\bf E}: 
	The afterglow emission (red) favors a parallel magnetic field ($\psi'< 30^\deg$) in the Envelope and X-ray data alone are sufficient for that conclusion 
	(green vs red). Adding the prompt emission (black) weakens that results. These conclusions stand even if the prompt emission were softer ($\bL = -1/3$ 
	for thin line, $\bL = 1/2$ for thick line) or if the emission-epoch $\to$ were free (or fixed to another epoch prior to the first XRT measurement) (issue \#1). 
	These results suggest a {\bf perpendicular $B$ in the Core} and a {\bf parallel $B$ in the Envelope}.
	Confidence in the former suggestion is 
	(1) strengthened by the $\Ep$ decreasing faster than the $LAE$ model expectation during the prompt phase (panel A), which would bias the best-fit 
	toward a parallel $B$ in the Core (issue \#2) and
	(2) weakened by the hardening of the XRT spectrum at the Tail end (panel B), which reduces the prompt light-curves curvature and mimicks the effect 
	of a perpendicular $B$ in the Core (issue \#5).
}
\end{figure}

\begin{figure}
\vspace*{-1cm}
\centerline{\includegraphics[width=15cm,height=16cm]{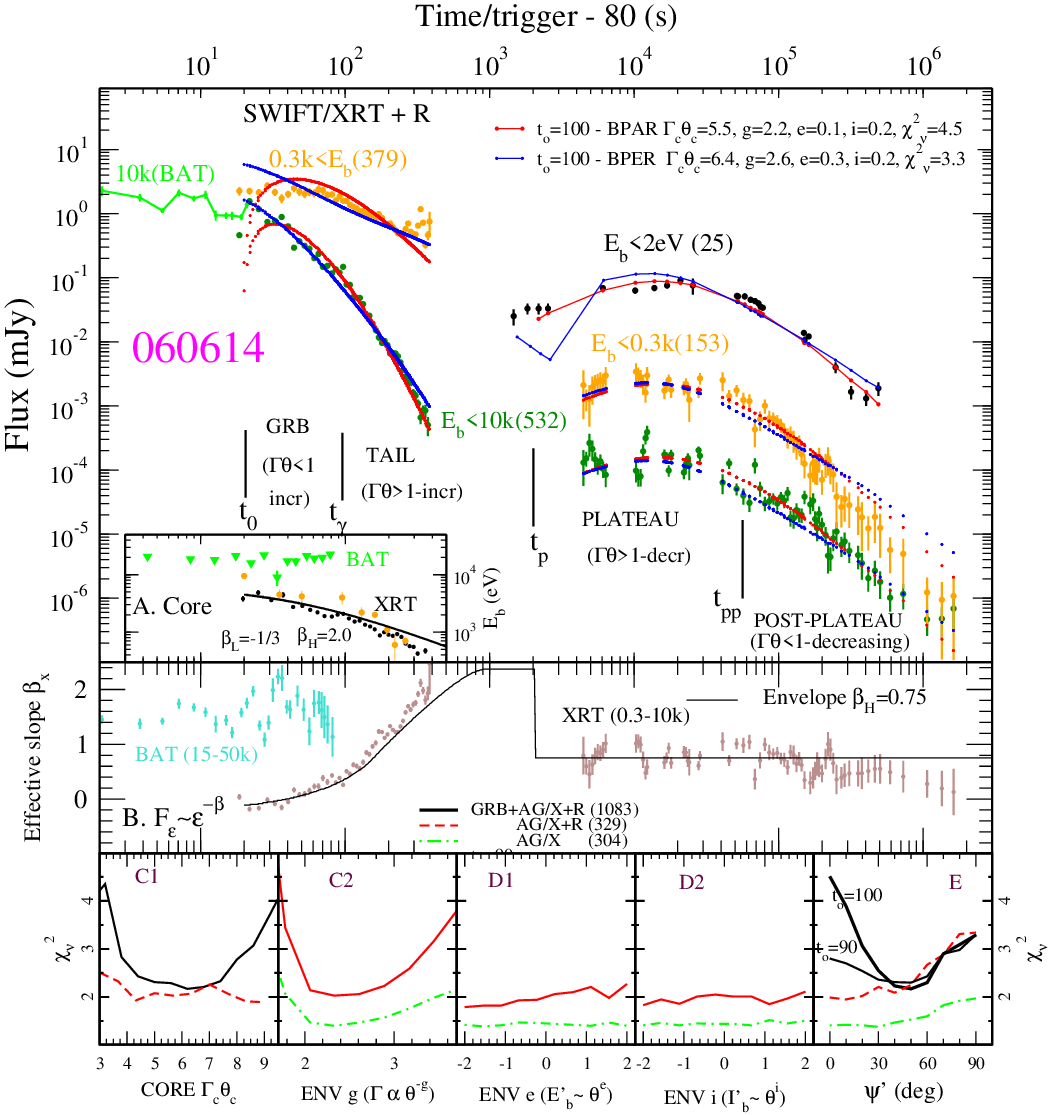}}
\figcaption{ {\bf GRB 060614}. 
Upper panel:
	The emission epoch is set at $\to=100$ s, at the first XRT measurement, but could be 10s earlier.
	{\bf The XRT has monitored most of the prompt phase}, the entire GRB and the first half of the Tail (in log space).
        Optical data from M. Della Vale, G. Chincarini \& N. Panagia (2006), J. Fynbo et al (2006), A. Gal-Yam et al (2006), and V. Mangano et al (2007).
Panel {\bf A}: 
	The BAT effective slope (panel B) does not display a systematic evolution, which indicates that the break-energy $\Ep$ is below 15 keV.
	Orange symbols show the $\Ep$ measurements of N. Butler \& D. Kocevski (2007) from direct spectral fits. 
	Our $\Ep$ determinations are within a factor 2 of those measurements. 
Panel {\bf B}:
        The BAT effective spectral slope (panel B) does not display a softening, thus the inferred $\Ep$ should be considered only upper limits.
	The XRT {\bf early asymptotic effective slope requires a low-energy spectral slope} $\bL=-1/3$.
	Because the Tail-to-Plateau transition was not monitored (upper panel), the last measured spectral slope $\bx$ for the prompt phase sets 
	only a hard limit $\bH=2.0$ on the Core's high-energy slope.
Panel {\bf C1}: 
	Because the Tail-to-Plateau transition was not monitored, the $\tp$ epoch is uncertain and the Core parameter $\Gctc (4.5,7.5)$
	cannot be determined, but just constrained.
Panel {\bf C2}: 
	The index for the Envelope's Lorentz factor, $g \in (2.0,2.9)$, is constrained mostly by the X-ray afterglow (green vs red).
Panels {\bf D1, D2}:
	The bright optical afterglow requires a break-energy $\Ep$ below optical (upper panel), and optical measurements do not constrain the afterglow asymptotic 
        flux-decay indices (they should be the same as in the X-ray), thus {\bf the four Envelope parameters should not be well constrained}.
	That is clearly illustrated by the structural indices the Envelope emission characteristics.
Panel {\bf E}: 
	The afterglow emission (red) favors a parallel magnetic field ($\psi'< 50^\deg$) in the Envelope, with the optical yielding a better constraint (green vs red). 
	This afterglow has the peculiar feature of a hardening of the post-Plateau emission, which increases the curvature of the 0.3 keV light-curve, which favors 
        a parallel $B$ in the Envelope.
	Adding the prompt emission (black) makes the case for a large angle $B$ in the Core, either in the form of the magic angle $\Psi'=55^\deg$,
         which is good proxy for an isotropic $B$, or in the form of a perpendicular $B$, with a later $\to=100$s (thick black line) strenghtening 
        that conclusion (issue \#1).
	The case for a perpendicular $B$ in the Core would be strengthened if the prompt emission were softer than $\bH = 2.0$ at the end of the Tail (issue \#4).
        Thus, this afterglow offers good evidence for and a {\bf parallel $B$ in the Envelope} and the prompt emission suggests a {\bf perpendicular/isotropic $B$ 
        in the Core}. 
}
\end{figure}

\begin{figure}
\vspace*{-1cm}
\centerline{\includegraphics[width=15cm,height=16cm]{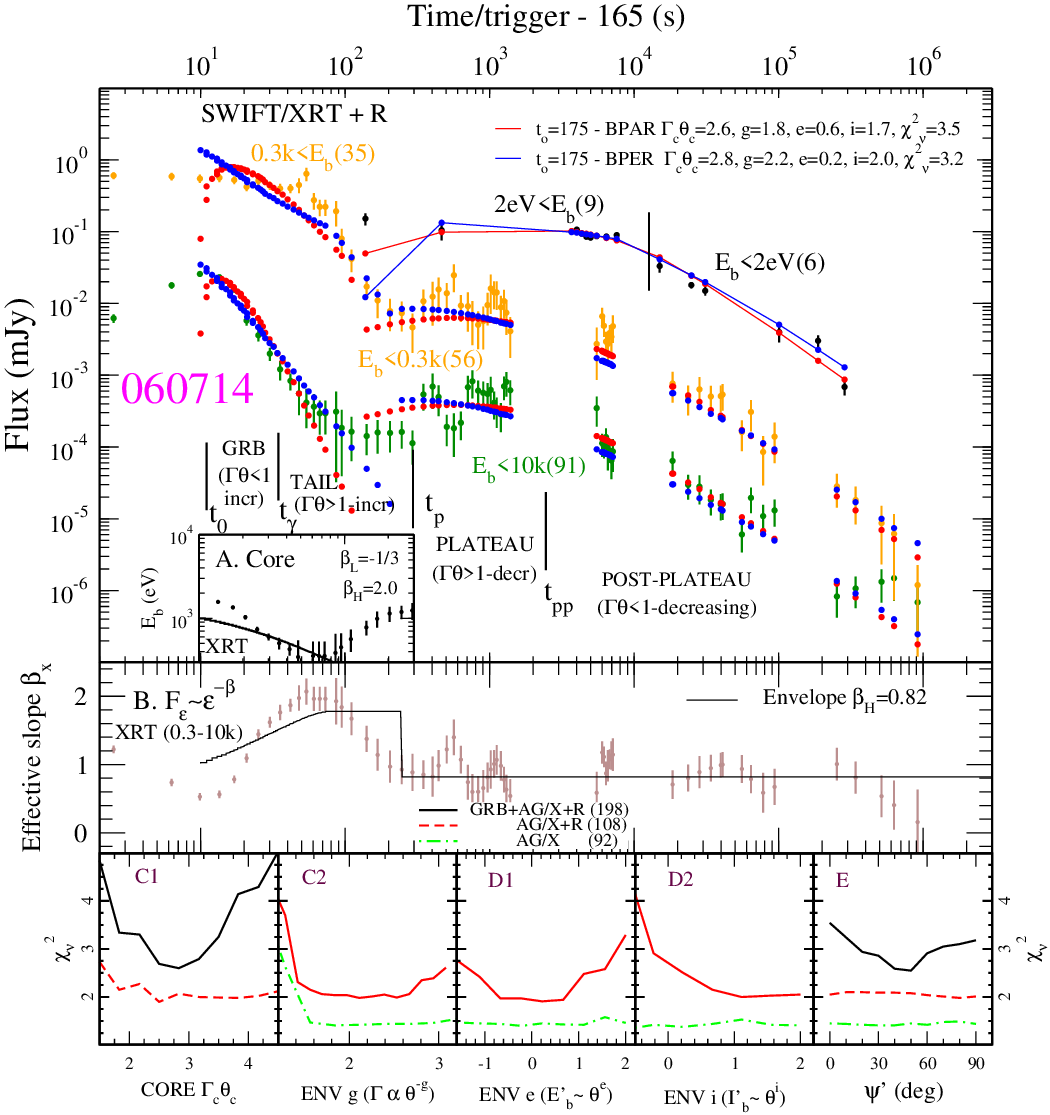}}
\figcaption{ {\bf GRB 060714}.
Upper panel:
        The emission epoch is at $\to=175$ s, at the peak of the third and last XRT pulse (the other two peaks, at 110-115 s and 130-140 s, are of comparable brightness).
        {\bf The XRT has monitored the entire prompt emission of this pulse}.
        Optical data from H. Krimm et al (2007).
Panel {\bf A}: 
        The measured break-energy $\Ep$ decreases faster than expected for the $LAE$ emission and $\Ep$ displays an increase toward the Tail end
        due to the hardening of the effective slope $\bx$ shown in panel B. 
Panel {\bf B}:
        The XRT {\bf early asymptotic effective slope $\bx$ allows for both low-energy spectral slopes} $\bL=-1/3$ and $\bL=1/2$.
        That effective XRT slope softens to $\bH=2.0$ during the Tail and hardens at the end of the Tail.
Panel {\bf C1}:
        The afterglow temporal milestones are not sufficiently well constrained, thus the Core parameter $\Gctc (2.4,3.2)$ is determined by the prompt emission.
Panel {\bf C2-E}: 
        The early (Plateau) asymptotic optical flux-decay is poorly constrained and the late (post-Plateau) optical flux-decay does not provide an new observational 
        constraint because the break-energy $\Ep$ is below optical for most of the post-Plateau, thus the afterglow data misses two observational constraints and
        the {\bf Envelope parameters should not be well constrained}: $g \in (1.5,2.7)$, $e \in (-0.8,0.8)$, $i > 2/3$.
Panel {\bf E}:
        As argued above, the afterglow emission (red) {\bf cannot determine the magnetic field orientation in the Envelope}. However, the prompt emission 
        (black line includes it) indicate an {\bf isotropic $B$ in the Core} (because $\chinu$ has a minimum in the vicinity of the magic angle $\Psi' = 55^\deg$, 
        with a best-fit bias for a parallel $B$ arising from that the model break-energy $\Ep$ decreases slower than measured (issue \#2) being compensated 
        by a bias for a perpendicular $B$ due to the spectral hardening at the Tail end (issue \#5). 
        These results stand even if the low-energy prompt emission were softer ($\bL=1/2$), which would favor a perpendicular $B$ (issue \#3) in the Core.
}
\end{figure}

\begin{figure}
\vspace*{-1cm}
\centerline{\includegraphics[width=15cm,height=16cm]{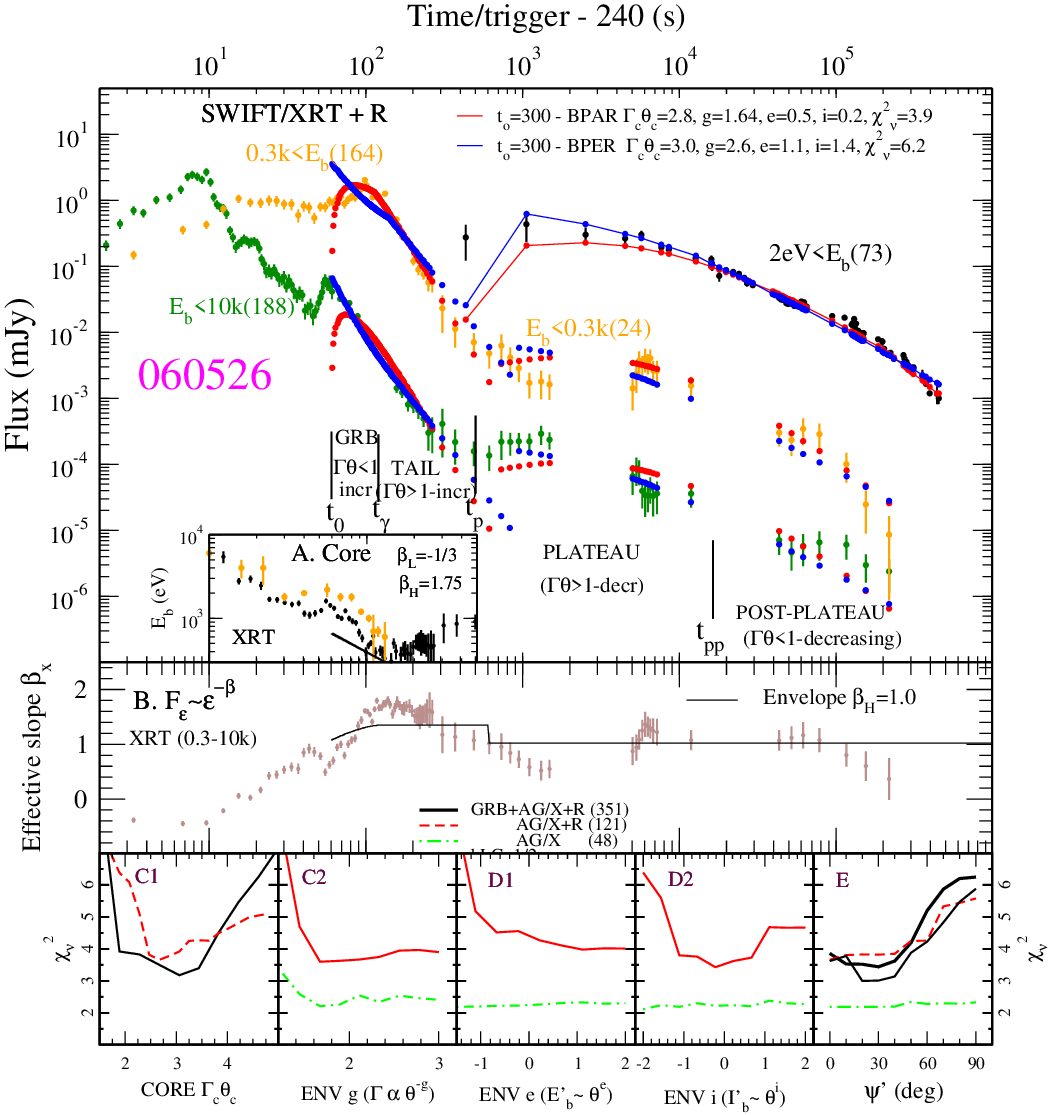}}
\figcaption{ {\bf GRB 060526}.
Upper panel:
        The emission epoch is set at $\to=300$ s, corresponding to the second, dimmer XRT pulse. 
        {\bf The XRT has monitored the entire prompt emission}.
        Fits to the prompt emission are much worse if $\to$ is set at 250s, the peak of the first XRT pulse because the current $LAE$ model with 
        a uniform Core cannot accommodate the long-lived flatness of the 0.3 keV light-curve (which is due to the softening of the XRT continuum - panel B).
        Optical data from X. Dai et al (2007) and C. Thone et al (2010).
Panel {\bf A}:
        Orange symbols show the break-energies $\Ep$ obtained by N. Butler \& D. Kocevski (2007) through spectral fits.
        Our values for $\Ep$, calculated from the XRT effective slope $\bx$, systematically underestimate those more accurate measurements 
        by a factor up to 2. 
        The measured break-energy $\Ep$ decreases faster than expected for the best-fit $LAE$ model.
Panel {\bf B}:
        For the entire XRT emission, the {\bf early asymptotic effective slope $\bx$ requires} $\bL=-1/3$, but the second peak's continuum allows 
        also for $\bL=1/2$. That effective XRT slope softens to $\bH=1.75$ during the Tail and hardens at the end of it.
Panel {\bf C1}:
        The prompt and afterglow temporal milestones seem sufficiently well-constrained (black vs red lines), leading to a Core parameter $\Gctc (2.7,3.5)$.
Panel {\bf C2-E}:
        The XRT afterglow coverage is somewhat sparse, thus the asymptotic power-law X-ray flux decays may not be well-constrained. 
        The break-energy $\Ep$ is between optical and X-ray for the entire afterglow, thus the optical data provide two observational constraints, 
        but, overall, the {\bf Envelope parameters should be insufficiently constrained}: $g \in (1.6,2.4)$, $e > 0$, $i \in (-1.0,0.8)$.
Panel {\bf E}:
       The sparser X-ray data (green) cannot constrain the $B$-field orientation in the Envelope, but the better-sampled optical afterglow light-curve 
       (included for red line) indicates {\bf a parallel or an isotropic field in the Envelope} (a low $\chinu (\psi)$ persists up to the magic angle 
       of $55^\deg$). Adding the prompt emission (thick black) suggests an isotropic $B$ in the Core (the dip in $\chinu (\psi')$ at $\psi'=20^\deg-40^\deg$). 
       The slower-than-measured decrease of the model break-energy $\Ep$ yields a bias toward a parallel $B$ (issue \#2) in the Core,
       which is compensated for the bias toward a perpendicular $B$ associated with the spectral hardening of the Tail end (issue \#5). 
       Thus, the evidence for an {\bf isotropic $B$ in the Core} appears reliable.  
       As expected, a softer low-energy slope of the prompt emission ($\bL=1/2$ -- thin line, $\bL=-1/3$ -- thick line) should introduce a bias toward
       a perpendicular $B$ (issue \#3).
}
\end{figure}

\begin{figure}
\vspace*{-1cm}
\centerline{\includegraphics[width=15cm,height=16cm]{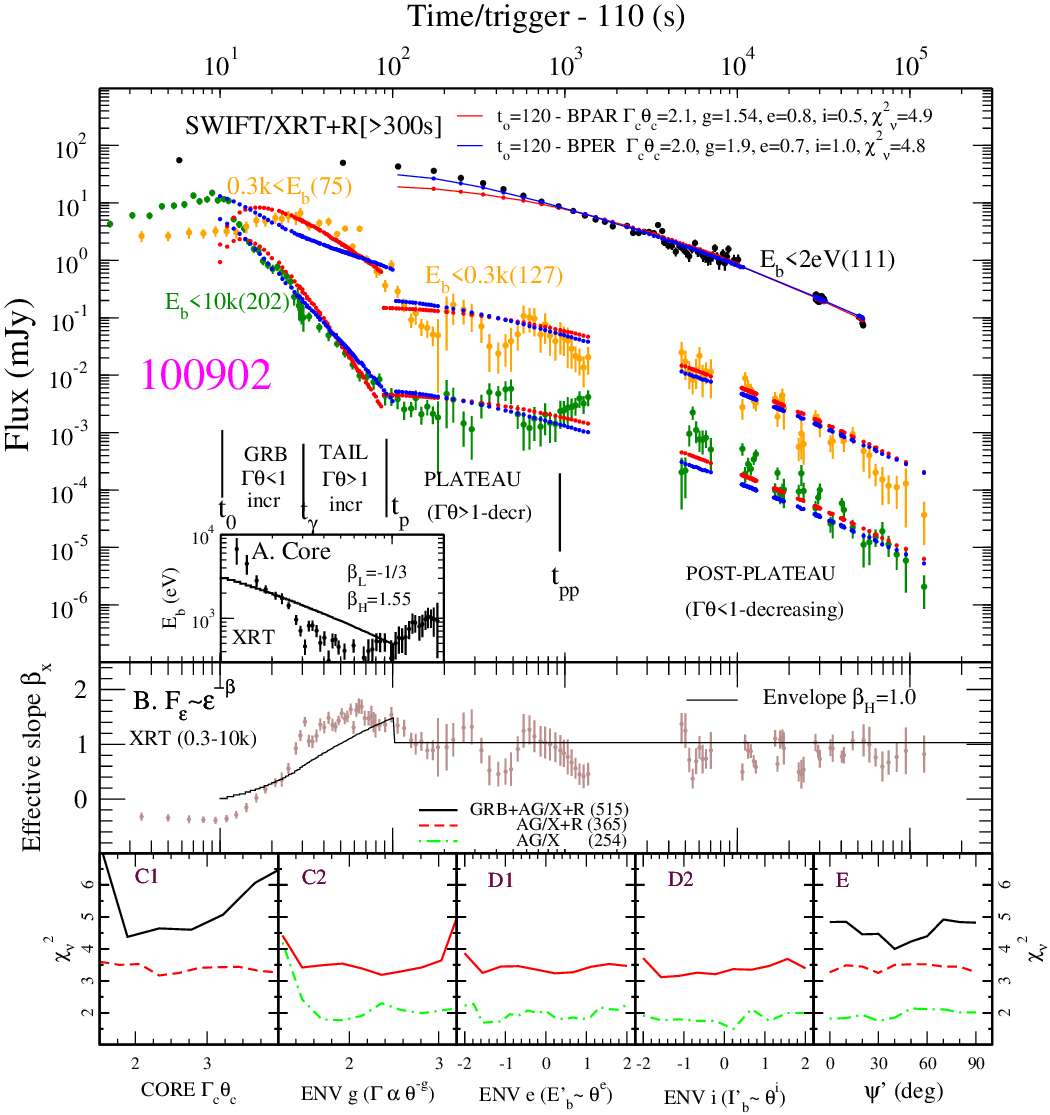}}
\figcaption{ {\bf GRB 100902}.
Upper panel:
        The emission epoch is at $\to=120$ s, at the only XRT pulse.
        {\bf The XRT has monitored the entire prompt emission}. The X-ray Plateau displays large flux fluctuations.
        The optical and X-ray afterglow light-curves are well-coupled, displaying similar asymptotic flux-decay indices.
        Optical data from E. Gorbovskoy et al (2011).
Panel {\bf A}:
        The measured break-energy $\Ep$ decreases significantly faster than expected for the $LAE$ emission.
Panel {\bf B}:
        The {\bf early asymptotic effective slope $\bx$ requires} $\bL=-1/3$.
        There is some hardening of the Tail end spectrum.
Panel {\bf C1}:
        The afterglow temporal milestones seem insufficiently constrained, perhaps due to the large fluctuations of the X-ray Plateau, and only the prompt emission
        constrains the Core parameter $\Gctc (2.0,3.0)$ (black vs red).
Panel {\bf C2-E}:
        For an Envelope emission spectral slope $\bH=1.0$ (panel B), the brightness of the optical afterglow requires that the break-energy $\Ep$ of the Envelope 
        emission is below optical, which implies that the optical does not provide an independent measurement of the asymptotic flux decay indices of the afterglow
        light-curves. Without those indices, the {\bf afterglow emission cannot constrain well the Envelope parameters}: $g \in (1.4,2.9)$, $e > -1.6$ and 
        $i \in (-1.7,1.2)$.
Panel {\bf E}:
       The afterglow data {\bf do not constrain the $B$ field orientation in the Envelope} (red and green) and the prompt measurements (black) indicate
       an {\bf isotropic $B$ in the Core} because the dip in $\chinu (\psi')$ at $\psi'=40^\deg$ is close to the magic angle of $55^\deg$).
       The bias for a parallel $B$ in the Core induced by slower-than-measured decrease of the model break-energy $\Ep$ (issue \#2)
       is compensated for the bias toward a perpendicular $B$ associated with the spectral hardening of the Tail end (issue \#5).
}
\end{figure}

\begin{figure}
\vspace*{-1cm}
\centerline{\includegraphics[width=15cm,height=16cm]{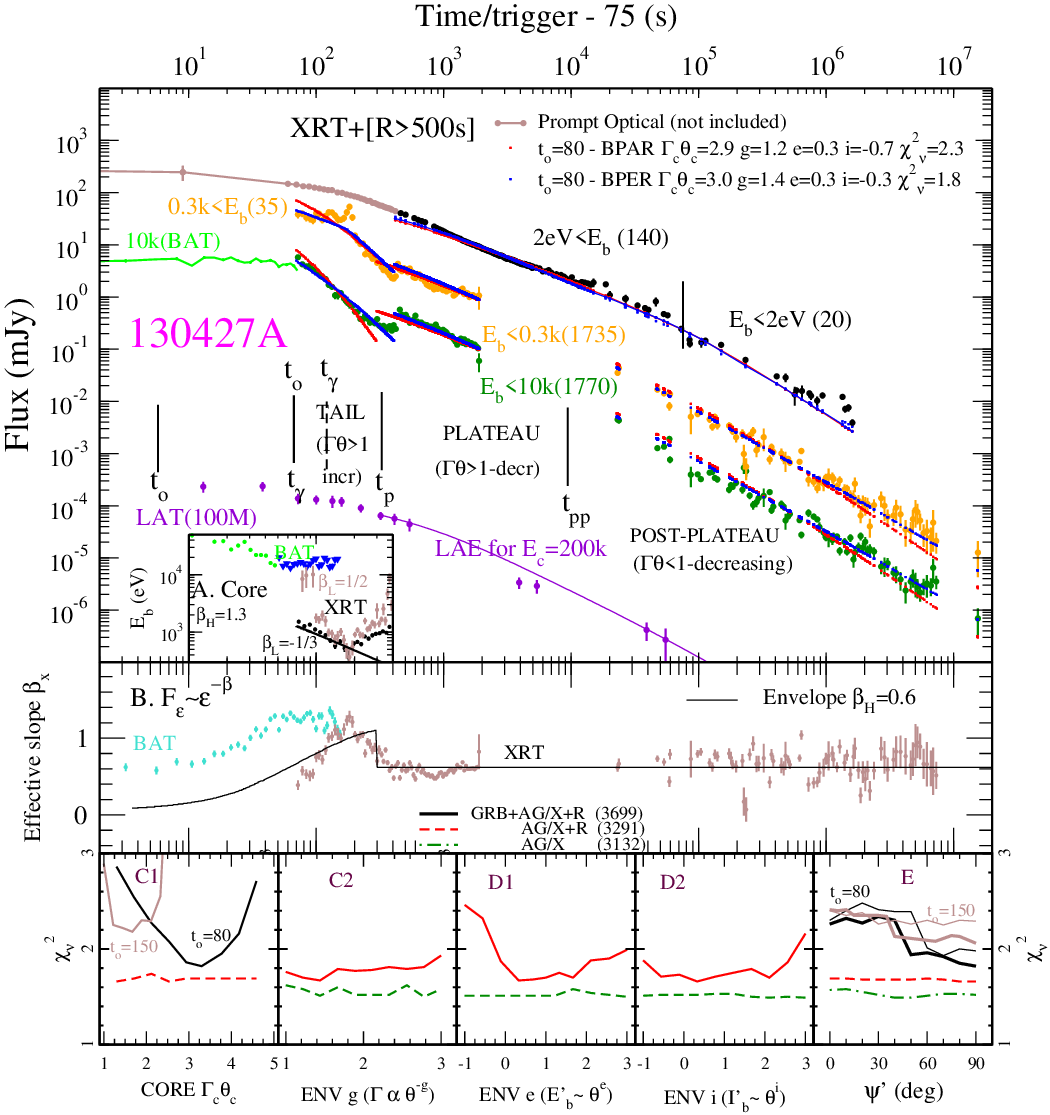}}
\figcaption{ {\bf GRB 130427}.
Upper panel: 
        The $LAE$ emission-epoch is at $\to=80-150$ s, within the broad peak of the BAT 10 keV light-curve and up to the beginning of the XRT coverage.
	This implies that the XRT may have missed the GRB phase and {\bf has monitored most of the Tail}.
        Only afterglow optical data after 500 s are included in the fit because the prompt optical emission does not show a Tail and
	appears decoupled from the X-ray. 
	Prompt LAT measurements are not included in the fit because they also do not display a Tail and afterglow LAT data are excluded 
	because the assumption of a single power-law spectrum from $\Ep \sim 1k$ to 100 MeV, with the best-fit high-energy slope $\bH = 0.6$, 
	over-produces 100 MeV photons. Such a hard spectrum implies an increasing emission output with photon energy, each higher decade 
	containing 2.5 times more energ than the next lower decade. Both of these point to the likely existence of a higher break-energy. 
	The 100 MeV model light-curve shown corresponds to a fixed break being located at $\Ec = 200$ keV and with a spectral softening 
	$\delta \bH = 1/2$ across it, as expected for the cooling-break. 
        Optical data from D. Perley et al (2014) and T. Vestrand et al (2014).
Panel {\bf A}: 
        The spectral softening shown in panel B leads to decreasing break-energies in the BAT and XRT domains.
        The BAT effective slope asymptotes to a constant values after 125 s, indicating that the break-energy $\Ep$ has fallen below 15 keV
        thus the subsequent values (blue) are upper limits. The XRT break-energy $\Ep$ is shown for two values of the low-energy slope $\bL=-1/3$ (black)
        and $\bL=1/2$ (brown) of panel B. For $\bL=1/2$, the break-energy makes a continuous but fast decay from the BAT to the XRT range.
        For $\bL=-1/3$, that transition is a sudden decrease. 
Panel {\bf B}:
	The BAT (turquoise) early effective slope $\bx$ asymptotes at $\bL=1/2$ and the XRT early slope $\bx$ is {\bf consistent with a low-energy
        slope $\bL=-1/3$ and $\bL=1/2$}. For this burst, there is good evidence that the spectral slopes of the broken power-law spectrum
        are the same for BAT and XRT. The Tail end displays a hardening of the high-energy slope $\bH$.
Panel {\bf C1}: 
	That the Core parameter $\Gctc$ is not determined by the afterglow light-curves (red) indicates that the Plateau to post-Plateau  
        epoch $\tpp$ is not well constrained by the optical light-curve (big gap around $\to$ in the XRT coverage).
	The uncertainty in the emission-epoch $\to$ leads to a broad range of $\Gctc \in (1.2,3.9)$ constrained by the prompt emission (black).
Panels {\bf C2-E}:
       The break-energy $\Ep$ is between optical and X-ray mostly during the Plateau, and one observational constraint (the post-Plateau asymptotic
       flux-decay indices) is missing, thus the {\bf afterglow data may not constrain well all Envelope parameters}: $g \in (1,3)$, $e \in (0.2,2.0)$,
        and $i \in (-0.8,2.4)$.
Panel {\bf E}: 
       The afterglow emission (red) cannot {\bf constrain the magnetic field orientation in the Envelope} because of the lack of one observational 
       constraint (above) and because the light-curve curvature associated with $\psi'$ can be easily mimicked by the other uncertain Envelope parameters 
       ($g;e,i$).
       Adding the prompt emission (black) favors a perpendicular $B$ in the Core for $\to=80$ s (black) and for either low-energy slope of 
       the prompt emission ($\bL=-1/3$ -- thick line, $\bL=1/2$ -- thin line). The same is true for a later $\to=150$ s (brown) if $\bL=-1/3$, 
       but evidence for that $B$ orientation evaporates if $\bL=1/2$ (thin line), despite that a softer low-energy slope $\bL$ of the Core emission 
       should bias the best-fit toward a perpendicular $B$ (issue \#3).
       This conclusion is weakend by the spectral hardening at the Tail end, which biases the best-fits toward a perpendicular $B$ (issue \#5).
       A later emission-epoch $\to < 150$s does not favor more a perpendicular field (issue \#1) because there is no XRT coverage prior to 150s. 
       Thus, the prompt emission of this burst provides some evidence for a {\bf perpendicular $B$ in the Core}. 
}
\end{figure}


\end{document}